\documentclass[aps, pra, showpacs,superscriptaddress , unsortedaddress, twocolumn]{revtex4-1}
\usepackage{graphicx}
\usepackage{calc}
\usepackage{amsmath}
\usepackage{amssymb}
\usepackage[english]{babel}

\begin{document}

\title{Theory of Elastic Interaction of the Colloidal Particles in the Nematic Liquid Crystal Near One Wall and in the Nematic Cell}

\author{S. B. Chernyshuk and B. I. Lev }

\affiliation{ Institute of Physics, NAS Ukraine, Prospekt
Nauki 46, Kyiv 03650, Ukraine}

\affiliation{ Bogolyubov Institute of
Theoretical Physics,NAS Ukraine,Metrologichna 14-b, Kyiv 03680,
Ukraine.}

\date{\today}

\begin{abstract}
We apply the method developed in Ref. [S.B.Chernyshuk and B.I.Lev, Phys.Rev.E, \textbf{81}, 041701 (2010)] for theoretical investigation of colloidal elastic interactions between axially symmetric particles in the confined nematic liquid crystal (NLC) near one wall and in the nematic cell with thickness $L$. Both cases of homeotropic and planar director orientations are considered. Particularly dipole-dipole, dipole-quadrupole and quadrupole-quadrupole interactions of the \textit{one} particle with the wall and within the nematic cell are found as well as corresponding \textit{two particle} elastic interactions.
A set of new results has been predicted:  the effective power of repulsion between two dipole particles at height $h$ near the homeotropic wall is reduced gradually from inverse 3 to 5 with an increase of dimensionless distance $r/h$; near the planar wall - the effect of dipole-dipole \textit{isotropic attraction} is predicted for large distances $r>r_{dd}=4.76 h$; maps of attraction and repulsion zones are crucially changed for all interactions near the planar wall and in the planar cell; one dipole particle in the homeotropic nematic cell was found to be shifted by the distance $\delta_{eq}$ from the center of the cell \textit{independent} of the thickness $L$ of the cell. 
The proposed theory fits very well with experimental data for the confinement effect of elastic interaction between spheres in the homeotropic cell taken from [M.Vilfan et al. Phys.Rev.Lett. {\bf 101}, 237801, (2008)] in the range $1\div1000 kT$.

\end{abstract}


\maketitle

\section{Introduction}
Colloidal particles in nematic liquid crystals (NLC) have attracted a large amount of research interest over the last few years. Anisotropic properties of the host fluid - liquid crystal - give rise to a new class of colloidal anisotropic interactions that never occur in isotropic hosts. The world of anisotropic  liquid crystal colloids is much more varied than that of isotropic liquids. 

Experimental study of anisotropic colloidal interactions in the bulk NLC has been made in \cite{po1}-\cite{jap2}. These interactions result in different structures of colloidal particles, such as linear chains along the director field, for particles with dipole symmetry of the director, \cite{po1,po2} and inclined chains with respect to the director for quadrupole particles \cite{po2}-\cite{kot}. Colloidal particles suspended at the nematic-air interface form 2D hexagonal structures \cite{nych,R10}. Quasi two-dimensional nematic colloids in thin nematic cells form a rich variety of 2D crystals by using laser tweezers. There are 2D hexagonal quadrupole crystals \cite{Mus,s1}, anti-ferroelectric dipole type 2D crystals \cite{Mus,s2} and mixed 2D crystals \cite{ulyana} sandwiched between cell walls. Levitation effect of dipole particles in planar cells was studied in \cite{lavr3}.
Long ranged elastic interactions between colloids have been experimentally found to be exponentially screened (confinement effect) in the nematic cell across distances compared to the cell thickness $L$ \cite{conf}.
Experimental results are reproduced by using the Landau - de Gennes free-energy numerical minimization approach \cite{Mus},\cite{conf}-\cite{stark2} as well as molecular dynamics \cite{andr1}. 

Theoretical understanding of the matter in the bulk NLC is based on the multipole expansion of the director field and has deep electrostatic analogies \cite{lupe}-\cite{perg3}. In spite of some differences in these approaches, only one of them \cite{lupe} gives an exact quantitative result which has been proven experimentally. Authors of \cite{noel} measured directly the dipole-dipole interaction of iron spherical particles in a magnetic field and found it to be in accordance with \cite{lupe}, within a few percents accuracy for a dipole term. Authors of \cite{jap} and \cite{jap2} have measured experimentally dipole-dipole interaction and found it to be in accordance with  \cite{lupe} within about $10\%$ accuracy. This allows to justify main assumptions of \cite{lupe} for spherical particles in infinite nematic liquid crystal. 

In spite of some understanding of the matter in the bulk NLC, there was no theoretical approach which was able to make exact quantitative predictions for nematic colloids in the confined liquid crystal, though almost all experiments in the nematic liquid crystal field are being conducted exactly in the confined volumes like walls of interface, cells \textit{etc}. In papers \cite{fukuda1,fukuda}, authors explained qualitatively screening effects within the \textit{coat} approach \cite{lev3} developed for the case of the homeotropic cell. But the coat approach can not give exact quantitative results for potentials as the parameters of the coat remain unknown. 

Recently, a proposal was made to take another approach \cite{we} for the description of colloidal particles in the confined NLC, which may be a generalization of the method \cite{lupe} for confined NLC. Using this approach, it is possible to find all long-range asymptotic behavior of the colloids in confined NLC and to make exact quantitative predictions which can be tested and compared with the experiment.

 In this paper we apply the proposed Green function method \cite{we} for quantitative description of the elastic colloidal interactions between axially symmetrical colloidal particles near one wall and in the nematic cell. We predict many new effects that fit very well with experimental data for the confinement effect of elastic interaction between spheres in the homeotropic cell taken from \cite{conf} in the range $1\div1000 kT$. 
 
The outline of the paper is the following : In Sec. II we talk about an exact approach for colloidal particles in nematic liquid crystals and topological defects. Sec. III presents the general Green function method for the description of colloidal particles in confined NLC, Sec. IV presents the energy of \textit{one} particle near one wall with both homeotropic and planar anchoring conditions, in Sec. V and Sec. VI we present the energy of elastic interaction between \textit{two} particles near one wall with homeotropic and planar anchoring conditions, respectively. Sec. VII presents the energy of \textit{one} particle in the NLC cell with a thickness $L$ with homeotropic and planar anchoring conditions. In Sec. VIII we find interactions between \textit{two} particles within a homeotropic nematic cell and in Sec. IX interactions between \textit{two} particles within a planar nematic cell are found. In Sec. X we make an analysis of the obtained results and discuss them in comparison with the results of other authors . And in Sec. XI we make our conclusions.

\section{The exact approach for colloidal particles in nematic liquid crystals. Topological defects.}

In this section we show the exact general formulation of the problem for a system of colloidal particles in nematic liquid crystal.

 Let's consider first one colloidal particle of the size $0.1\div 10 \mu m $ inside the unlimited nematic liquid crystal. The free energy of the system consists of the deformational energy of the LC and surface energy of the particle. Deformational energy can be written in the well known Frank form:
\begin{widetext}
\begin{equation}
F_{bulk}=\frac{1}{2}\int d^{3}x\left\{K_{1}(\nabla\cdot\textbf{n})^{2}+K_{2}(\textbf{n}\cdot\nabla\times\textbf{n})^{2}+K_{2}(\textbf{n}\times\nabla\times\textbf{n})^{2}\right\}\label{fb}
\end{equation}
\end{widetext}
where $\textbf{n}$ is the unit vector pointing average orientation of the long axes of LC molecules (we don't take into account $K_{24}$ and $K_{13}$ terms for simplicity, though they exist in the full deformation energy. Full energy derivation of NLC with $K_{24}$ and  $K_{13}$ terms from microscopic theory was obtained in \cite{pergchern}). In the one constant approximation $K=K_{ii}$ the total bulk deformation energy has the form:
\begin{equation}
F_{bulk}=\frac{K}{2}\int  d^{3}x \left[(\nabla\cdot\textbf{n})^{2}+(\nabla\times\textbf{n})^{2}\right] \label{fb1}
\end{equation}
The magnitude of the elastic constant is $K\approx 10 pN$.

On the surface of the particle, LC molecules have the tendency to lie either perpendicular or parallel to the surface. The orientation depends on the coating of the surface. The resulting surface energy can be written in the Rapini-Popula form:
\begin{equation}
F_{surface}=W\oint d\sigma(\textbf{n}\nu )^{2}\label{fs}
\end{equation}
with $\nu$ normal vector to the surface, $W<0$ corresponds to the normal orientation of LC molecues (homeotropic anchoring) and $W>0$ corresponds to the parallel orientation of LC molecues (planar anchoring).  Typical scale of the anchoring coefficient is $W=5\cdot10^{-5}J/m^{2}$. Just this surface term is the origin of bulk director deformations.  For small particles with size $R<r_{c}=\frac{K}{W}\approx 0.2\mu m$ deformations are small but when the size of the particle exceeds $R>r_{c}=\frac{K}{W}\approx 0.2 \mu m$ ,surface energy plays a dominant role and the director field follows the surface. Naturally, topological defects appear near the particle.  Fig.1 demonstrates examples of topological defects near the spherical particle \cite{stark1}. 

\begin{figure}[ht!]
\includegraphics[clip=,width=\linewidth]{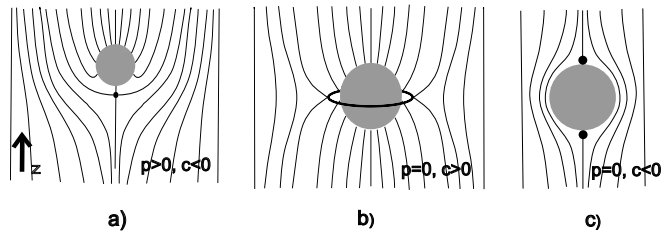}
\caption{\label{defects} Topological defects. a) Dipole configuration. Point defect - hyperbolic hedgehog, b) Saturn ring quadrupole configuration with $c>0$ and c) Two point defects - boojums at the poles of the sphere, quadrupole configuration with $c<0$.}
\end{figure}

Symmetry of the defects can be either dipole (case a.) or quadrupole (cases b. and c. ) for large distances. Such topological defects can not be treated precisely analytically, but, rather, with the help of approximate variational ansatzes or with the help of numerical simulations. 

If we have a system of many colloidal particles we need to replace (\ref{fs}) with:
\begin{equation}
F_{surface}=\sum_{i}W_{i}\oint_{i} d\sigma(\textbf{n}\nu )^{2}\label{fs2}
\end{equation}
where $i$ enumerate particles and to find the global minimum of the functional $F=F_{bulk}+F_{surface}$.  Moreover if we confine the liquid crystal with some bounding walls we need to add surface energy at each interface:
\begin{equation}
F_{walls}=\sum_{\gamma}\oint_{\Sigma_{\gamma}} W_{\gamma} d\sigma(\textbf{n}\nu )^{2}\label{fs3}
\end{equation}
where $\Sigma_{\gamma}$ are confining walls. Then we need to find the global minimum of the total free energy $F=F_{bulk}+F_{surface}+F_{walls}$. Since the problem cannot be solved precisely, even for one particle, without confining walls; it is much more difficult to solve it for a system of many particles that are found in the presence of bounding walls. 

Therefore it is necessary to introduce some another approach which can effectively describe the system of colloidal particles in the confined nematic liquid crystal. This analytical approach we will develop in the next section.

\section{General Green Function Approach for the Description of Colloidal Particles in Confined NLC}
Here we will describe, briefly, the method proposed in \cite{we}. Consider axially symmetric particle of the micron and sub-micron size which may carry topological defects such as hyperbolic hedgehog, disclination ring or boojums. In the absence of the particle ground, the non-deformed state of NLC is the orientation of the director $\textbf{n}|| z, \textbf{n}=(0,0,1)$. The immersed particle induces deformations of the director in the perpendicular directions $n_{\mu}, \mu=x,y$. In close vicinity to the particle, strong deformations and topological defects arise, but beyond them, deformations become small.
In the paper \cite{lev3} this area with strong deformation and defects was called the \textit{coat}. Beyond the coat, the bulk energy of deformation may be approximately written in the form:
\begin{equation}
\tilde{F}_{bulk}=\frac{K}{2}\int  d^{3}x (\nabla n_{\mu})^{2} \label{fb3}
\end{equation}
with Euler-Lagrange equations of Laplace type:
\begin{equation}
\Delta n_{\mu}=0  \label{lap}
\end{equation}
Then director field outside the coat in the infinite LC has the form $n_{x}(\textbf{r})=p\frac{x}{r^{3}}+3c\frac{xz}{r^{5}},n_{y}(\textbf{r})=p\frac{y}{r^{3}}+3c\frac{yz}{r^{5}}$ with $p$ and $c$ being dipole and quadrupole elastic moments (we use another notation for $c$ with respect to the $\tilde{c}$ in \cite{lupe}, so that our $c=\frac{2}{3}\tilde{c}$ ).
Actually $p$ and $c$ are unknown quantities. They can be found only as asymptotics from exact solutions or from variational ansatzes. It was found in \cite{lupe} that $p=\alpha a^2$, $c=-\beta a^3$ with $a$ being the particle radius, and for instance $\alpha=2.04$, $\beta=0.72$ for hyperbolic hedgehog configuration (see Fig.\ref{defects} a ) from ansatz \cite{lupe}. Experiment \cite{noel} gives $\alpha=2.05$, $\beta=0.2\pm0.1$, experiment \cite{jap} gives $\alpha=1.88\pm0.18$, $\beta=0.52\pm0.12$, experiment \cite{jap2} gives $\alpha=2.21\pm0.04$, $\beta=0.497\pm0.09$. Consider that we have found $p$ and $c$ and they \textit{are fixed}. Then we can formulate an effective theory which describes particles and interactions between them, remarkably well. In other words, constants $p$ and $c$ \textit{are bridges} between the effective theory and exact theory with free energy (\ref{fb1}) and (\ref{fs}).

 In order to find effective energy of the system: particle(s) + LC it is necessary to introduce some effective functional $F_{eff}$ so that it's Euler-Lagrange equations would have the above solutions. In the \cite{lupe} it was found that in the one constant approximation with Frank constant $K$, the effective functional has the form:
\begin{widetext}
\begin{equation}
F_{eff}=K\int d^{3}x\left\{\frac{(\nabla n_{\mu})^{2}}{2}-4\pi P(\textbf{x})\partial_{\mu}n_{\mu}-4\pi C(\textbf{x})\partial_{z}\partial_{\mu}n_{\mu} \right\}\label{flin}
\end{equation}
\end{widetext}
which brings Euler-Lagrange equations:
\begin{equation}
\Delta n_{\mu}=4\pi\left[\partial_{\mu}P(\textbf{x})-\partial_{z}\partial_{\mu}C(\textbf{x})\right] \label{nmu}
\end{equation}
where $P(\textbf{x})$ and $C(\textbf{x})$ are dipole and quadrupole moment densities, $\mu=x,y$ and repeated $\mu$ means summation on $x$ and $y$ like $\partial_{\mu}n_{\mu}=\partial_{x}n_{x}+\partial_{y}n_{y}$.
For the infinite space the solution has the known form: $n_{\mu}(\textbf{x})=\int d^{3}\textbf{x}' \frac{1}{\left|\textbf{x}-\textbf{x}'\right|}\left[ -\partial_{\mu}'P(\textbf{x}')+\partial_{\mu}'\partial_{z}'C(\textbf{x}') \right] $. If we consider $P(\textbf{x})=p\delta(\textbf{x})$ and $C(\textbf{x})=c\delta(\textbf{x})$ this really brings $n_{x}(\textbf{r})=p\frac{x}{r^{3}}+3c\frac{xz}{r^{5}},n_{y}(\textbf{r})=p\frac{y}{r^{3}}+3c\frac{yz}{r^{5}}$.
This means that effective functional (\ref{flin}) correctly describes the interaction between particle and LC via linear terms $-4\pi P(\textbf{x})\partial_{\mu}n_{\mu}-4\pi C(\textbf{x})\partial_{z}\partial_{\mu}n_{\mu}$. So it can be used, as well, for the description of colloidal particles in confined nematic liquid crustals.

In the case of confined nematic LC with the boundary conditions $n_{\mu}(\textbf{s})=0$ on the surfaces $\Sigma$ (Dirichlet boundary conditions)
the solution of EL equation (\ref{nmu}) has the form:
\begin{equation}
n_{\mu}(\textbf{x})=\int_{V} d^{3}\textbf{x}' G(\textbf{x},\textbf{x}')\left[ -\partial_{\mu}'P(\textbf{x}')+\partial_{\mu}'\partial_{z}'C(\textbf{x}') \right] \label{solmain}
\end{equation}
where $G$ is the Green function $\Delta_{\textbf{x}}G(\textbf{x},\textbf{x}')=-4\pi \delta(\textbf{x}-\textbf{x}')$ for  $\textbf{x},\textbf{x}'\in \textbf{V}$ ($\textbf{V}$ is the volume of the bulk NLC) and $G(\textbf{x},\textbf{s})=0 $ for any \textbf{s} of the bounding surfaces $\Sigma$. The mathematical symmetry property $G(\textbf{x},\textbf{x}')=G(\textbf{x}',\textbf{x})$ can be proved for the Green functions satisfying the Dirichtle boundary conditions by means of Green's theorem \cite{jac}.

 Consider $N$ particles in the confined NLC, so that $P(\textbf{x})=\sum_{i}p_{i}\delta(\textbf{x}-\textbf{x}_{i})$ and $C(\textbf{x})=\sum_{i}c_{i}\delta(\textbf{x}-\textbf{x}_{i})$. Then substitution (\ref{solmain}) into $F_{eff}$ brings: $F_{eff}=U^{self}+U^{interaction}$ where  $ U^{self}=\sum_{i}U_{i}^{self} $ , here $U_{i}^{self}$ is the interaction of the $i$-th particle with the bounding surfaces $ U_{i}^{self}=U_{dd}^{self}+U_{dQ}^{self}+U_{QQ}^{self} $. In the general case, the interaction of the particle with bounding surfaces, (self-energy part), takes the form:  
\begin{equation}
 U_{dd}^{self}=-2\pi K p^{2}\partial_{\mu}\partial_{\mu}'H(\textbf{x}_{i},\textbf{x}_{i}')|_{\textbf{x}_{i}=\textbf{x}_{i}'}\label{uself}
\end{equation}
$$
 U_{dQ}^{self}=-2\pi K pc( \partial_{\mu}\partial_{\mu}'\partial_{z}'H(\textbf{x}_{i},\textbf{x}_{i}')+\partial_{\mu}'\partial_{\mu}\partial_{z}H(\textbf{x}_{i},\textbf{x}_{i}'))|_{\textbf{x}_{i}=\textbf{x}_{i}'} 
$$
 $$
 U_{QQ}^{self}=-2\pi K c^{2} \partial_{z}\partial_{z}'\partial_{\mu}\partial_{\mu}'H(\textbf{x}_{i},\textbf{x}_{i}')|_{\textbf{x}_{i}=\textbf{x}_{i}'}
 $$
where  $G(\textbf{x},\textbf{x}')=\frac{1}{|\textbf{x}-\textbf{x}'|}+H(\textbf{x},\textbf{x}') $ and  $\Delta_{\textbf{x}}H(\textbf{x},\textbf{x}')=0$ (we excluded the divergent part of self energy from $\frac{1}{|\textbf{x}-\textbf{x}'|}$).\\
Interaction energy $ U^{interaction}=\sum_{i<j}U_{ij}^{int} $.  Here $U_{ij}^{int}$ is the interaction energy between $i$ and $j$ particles: 
 $ U_{ij}^{int}= U_{dd}+U_{dQ}+U_{QQ} $:
\begin{equation}
U_{dd}=-4\pi K pp'\partial_{\mu}\partial_{\mu}'G(\textbf{x}_{i},\textbf{x}_{j}') \label{uint}
\end{equation}

$$
U_{dQ}=-4\pi K \left\{pc'\partial_{\mu}\partial_{\mu}'\partial_{z}'G(\textbf{x}_{i},\textbf{x}_{j}')+p'c\partial_{\mu}'\partial_{\mu}\partial_{z}G(\textbf{x}_{i},\textbf{x}_{j}')\right\}
$$

$$
U_{QQ}=-4\pi K cc'\partial_{z}\partial_{z}'\partial_{\mu}\partial_{\mu}'G(\textbf{x}_{i},\textbf{x}_{j}')
$$
Here unprimed quantities are used for particle $i$ and primed for particle $j$. $U_{dd}, U_{dQ}, U_{QQ}$ means dipole-dipole, dipole-quadrupole and quadrupole-quadrupole interactions, respectively. 

Formulas (\ref{uself}) and (\ref{uint}) represent general expressions for the self energy of one particle, (energy of interaction with the walls), and interparticle elastic interactions in the arbitrary confined NLC with strong anchoring conditions $n_{\mu}(\textbf{s})=0$ on the bounding surfaces.

When bounding surfaces are located far away from the particles and their influence can be neglected, we have the approximation of unlimited nematic liquid crystal. In unlimited nematic $H(\textbf{x},\textbf{x}')=0$ and $G=G_{0}(\textbf{x},\textbf{x}')=\frac{1}{|\textbf{x}-\textbf{x}'|}$. Then self energy of the one particle is zero as we do not take into account divergent part of self energy from $G_{0}=\frac{1}{|\textbf{x}-\textbf{x}'|}$. As
well we do not take into account actual finite self energy (inside \textit{coat} region) of the one particle from the exact free energy (\ref{fb1}) and (\ref{fs}). Elastic interaction between particles in the infinite LC then has the well known form \cite{lupe}:
$$
\frac{U_{dd}}{4\pi K}=\frac{pp'}{r^{3}}(1-3cos^{2}\theta)
$$
\begin{equation}
\frac{U_{dQ}}{4\pi K}= (pc'-cp')\frac{cos\theta}{r^{4}}(15cos^{2}\theta-9)\label{udq}
\end{equation}
$$
\frac{U_{QQ}}{4\pi K}= \frac{cc'}{r^{5}}(9-90cos^{2}\theta+105cos^{4}\theta)
$$
where $r$ is the distance between particles, $\theta$ is the angle between $r$ and $z$.

\begin{figure}[ht!]
\includegraphics[clip=,width=\linewidth]{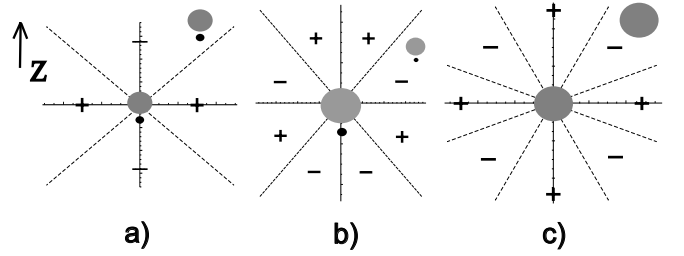}
\caption{\label{cones}Cones of repulsion and attraction for bulk interactions from (\ref{udq}), '-' means attraction, '+' means repulsion. a) Dipole-dipole interaction b) Dipole-quadrupole interaction with greater particle in the center c) Quadrupole-quadrupole interaction.}
\end{figure}

 Below we will apply the expressions (\ref{uself}), (\ref{uint}) for particular cases of the NLC confined with one wall as well as for NLC confined with two parallel walls (nematic cell) with homeotropic and planar boundary conditions . We shall consider below $p=\alpha a^2$, $c=-\beta a^3$ with $a$ being the particle radius.

\section{Interaction of the one particle with the wall}
\subsection{Interaction of the one particle with a homeotropic wall}
\begin{figure}[ht!]
\includegraphics[clip=,width=\linewidth]{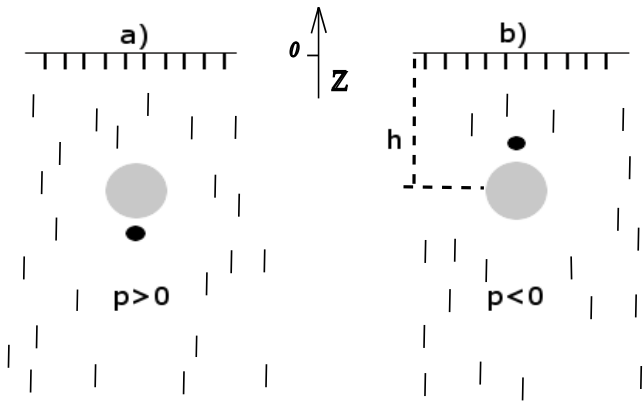}
\caption{\label{onehom}Repulsion of the one particle from the wall with homeotropic director orientation. a) \textit{p}-configuration b) \textit{h}-configuration. }
\end{figure}

We choose coordinate system for the wall with homeotropic conditions $z||\textbf{n}_{\infty}$, $z=0$ at the wall and $z>0$ above, so that particles and NLC under the wall have $z<0$ and heights $h=-z$ (see Fig.\ref{onehom} ). Then Green function in this case has the form \cite{jac} :
\begin{equation}
G_{hom}^{wall}(\textbf{x},\textbf{x}')=\frac{1}{|\textbf{x}-\textbf{x}'|}-\frac{1}{|\textbf{a}(\textbf{x})-\textbf{x}'|} \label{ghom}
\end{equation}

Here $\textbf{a}(\textbf{x})=(x,y,-z)$ for $\textbf{x}=(x,y,z)$. In other words $H(\textbf{x},\textbf{x}')=-\frac{1}{|\textbf{a}(\textbf{x})-\textbf{x}'|}$ is the $'potential'$ of the mirror image of the particle located in the point $\textbf{x}$ similar as in electrostatics. Then interaction of the particle with the wall consists of three parts $ U_{self}^{hom,wall}=U_{dd,self}^{hom,wall}+U_{dQ,self}^{hom,wall}+U_{QQ,self}^{hom,wall}$ and may be found via (\ref{uself}):
\begin{equation}
U_{dd,self}^{hom,wall}=\frac{\pi K \alpha^{2}a^{4}}{2h^{3}}\label{1}
\end{equation}
\begin{equation}
U_{dQ,self}^{hom,wall}=\mp\frac{3\pi K \alpha \beta a^{5}}{2h^{4}}\label{2}
\end{equation}
\begin{equation}
U_{QQ,self}^{hom,wall}=\frac{3\pi K \beta^{2}a^{6}}{2h^{5}}\label{3}
\end{equation}
where $h$ is the distance from the particle to the wall, '-' corresponds to $p>0$ (\textit{p}-configuration) and '+' corresponds to $p<0$ (\textit{h}-configuration, see Fig.\ref{onehom}). 
Formulas (\ref{1})-(\ref{3}) as well, may be obtained from Lubensky approach \cite{lupe} just considering that there is an image particle at the height $h$ above the wall with opposite dipole moment $p'=-p$ and the same quadrupole moment $c'=c$ and taking into account that interaction energy should be divided by two, as there is no real liquid crystal above the wall. For instance, energy of interaction between two dipole particles is $U_{dd}=\frac{4\pi Kpp'}{r^{3}}(1-3cos^{2}\theta)$, then taking $p'=-p=-\alpha a^{2},\theta=0,r=2h$ and divided by two we obtain formula (\ref{1}).
However this method cannot be applicable for the wall with planar orientation of the director. We shall demonstrate this in the subsection below.

\subsection{Interaction of the one particle with planar wall}
\begin{figure}[ht!]
\includegraphics[clip=,width=6.5cm, height=4cm]{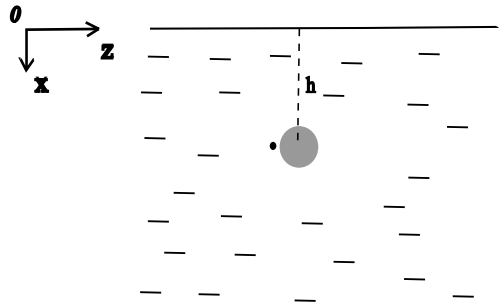}
\caption{\label{oneplan}Repulsion of the one particle from the wall with planar director orientation.}
\end{figure}

We choose coordinate system for the wall with planar conditions with $z||\textbf{n}_{\infty}$, axis $x$ looks down $0<x<\infty$, $x=0$ at the wall so that particles  and NLC have $x>0$ and heights $h=x$ (see Fig.\ref{oneplan}).

In order to find the Green function for this case, let's turn coordinate system (CS) of the homeotropic cell $CS^{hom}$ $(x,y,z)$ round the $y$ axis on $\pi/2$. Then we will have $CS^{plan}$ $(\tilde{x},\tilde{y},\tilde{z})$ with transition matrix $A$: $\textbf{x}=A\tilde{\textbf{x}},\textbf{x}'=A\tilde{\textbf{x}}'$ so that $x=\tilde{z},y=\tilde{y},z=-\tilde{x}$. Then $G_{hom}(\textbf{x},\textbf{x}')=G_{hom}(A\tilde{\textbf{x}},A\tilde{\textbf{x}}')=G_{plan}(\tilde{\textbf{x}},\tilde{\textbf{x}}')$. 

Omitting sign $\sim$ we may write the Green function for planar cell in the $CS^{plan}$ with $\textbf{n}||z$ and $x$ perpendicular to the cell plane ($x\in (0,\infty)$):
\begin{equation}
G_{plan}^{wall}(\textbf{x},\textbf{x}')=\frac{1}{|\textbf{x}-\textbf{x}'|}-\frac{1}{|\textbf{b}(\textbf{x})-\textbf{x}'|}\label{gplan}
\end{equation}
Here $\textbf{b}(\textbf{x})=(-x,y,z)$ for $\textbf{x}=(x,y,z)$. 
In other words $H(\textbf{x},\textbf{x}')=-\frac{1}{|\textbf{b}(\textbf{x})-\textbf{x}'|}$ is the $'potential'$ of the mirror image of the particle located in the point $\textbf{x}$. Taking derivatives of the $H(\textbf{x},\textbf{x}')$ we come to the interaction energy of the particle with the planar wall via (\ref{uself}): 
\begin{equation}
U_{dd,self}^{plan,wall}=\frac{3\pi K\alpha^{2}a^{4}}{4h^{3}} \label{1p}
\end{equation}
\begin{equation}
U_{dQ,self}^{plan,wall}=0
\end{equation}
\begin{equation}
U_{QQ,self}^{plan,wall}=\frac{15\pi K\beta^{2}a^{6}}{16h^{5}}\label{3p}
\end{equation}

We see that these results differ from (\ref{1})-(\ref{3}) for homeotropic wall. This means that planar orientation of the director on the wall violates the direct analogy between nematostatics and electrostatics, due to the different symmetry of the planar wall. Thus, interaction of the one particle with the planar wall cannot be treated using the previous results of \cite{lupe} as in the case of the homeotropic cell.

\section{Interaction between two particles near one homeotropic wall}
\begin{figure}[ht!]
\includegraphics[clip=,width=8cm, height=7cm]{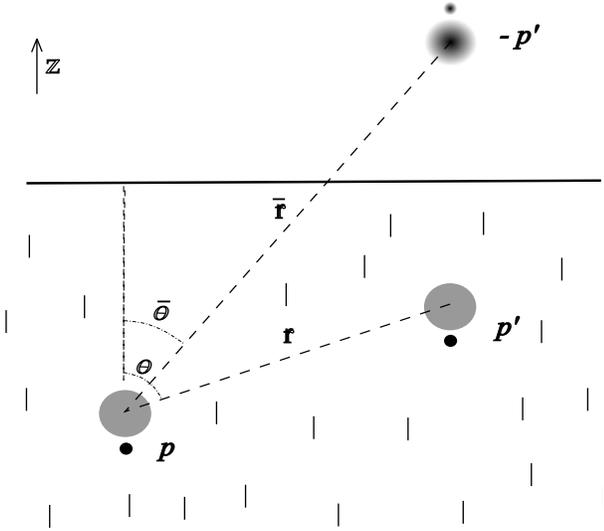}
\caption{\label{2hom}\textit{Image} interpretation of the interaction between two particles near the homeotropic wall. This barely supplements the exact formulas (\ref{uint}).}
\end{figure}

If we have the Green function (\ref{ghom}), we can find interaction between two particles using formulas (\ref{uint}).
Then taking derivatives, brings all necessary potentials of elastic interaction between two particles near one wall with homeotropic conditions:

\begin{widetext}
\begin{equation}
\frac{U_{dd,hom}^{wall}}{4\pi K}=\frac{pp'}{r^{3}}(1-3cos^{2}\theta)-\frac{pp'}{\bar{r}^{3}}(1-3cos^{2}\bar{\theta})\label{uddh}
\end{equation}
\begin{equation}
\frac{U_{dQ,hom}^{wall}}{4\pi K}= (pc'-cp')\frac{cos\theta}{r^{4}}(15cos^{2}\theta-9)+(pc'+cp')\frac{cos\bar{\theta}}{\bar{r}^{4}}(15cos^{2}\bar{\theta}-9)\label{udqh}
\end{equation}
\begin{equation}
\frac{U_{QQ,hom}^{wall}}{4\pi K}= \frac{cc'}{r^{5}}(9-90cos^{2}\theta+105cos^{4}\theta)+
\frac{cc'}{\bar{r}^{5}}(9-90cos^{2}\bar{\theta}+105cos^{4}\bar{\theta})
\label{uqqh}
\end{equation}
\end{widetext}
where $r$ is the distance between particles, $\theta$ is the angle between $r$ and $z$, $$\bar{r}=\sqrt{(x-x')^{2}+(y-y')^{2}+(z+z')^{2}}$$ is the distance between particle 1 and image of the particle 2 and $\bar{\theta}$ is the angle between $\bar{r}$ and vertical line (see Fig.\ref{2hom}). Then  $cos\bar{\theta}=-\frac{(z+z')}{\bar{r}}>0$ (we use coordinate system with $z||\textbf{n}$ and $z=0$ at the wall so that particles have $z<0$ under the wall). 

\begin{figure}[ht!]
\includegraphics[clip=,width=\linewidth]{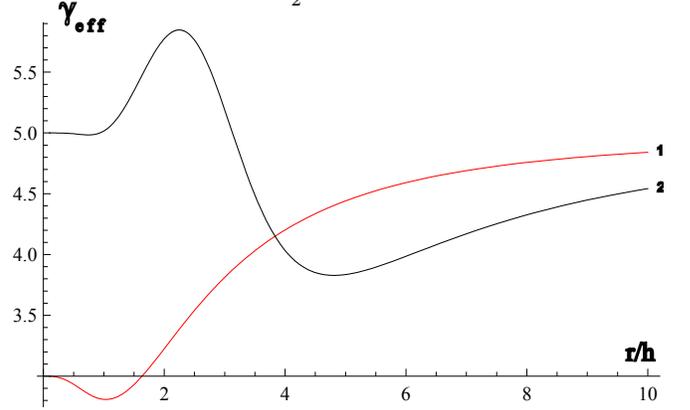}
\caption{\label{2power}(Color online) Effective power of repulsion between two particles at the same height $h$ near the homeotropic wall in (\ref{gamma}) vs. dimensionless distance $r/h$. Red line 1 corresponds to the dipole-dipole interaction, first term in (\ref{uhsame}); black line 2 - quadrupole-quadrupole interaction, third term in (\ref{uhsame}).}
\end{figure}

\begin{figure}[ht!]
\includegraphics[clip=,width=\linewidth]{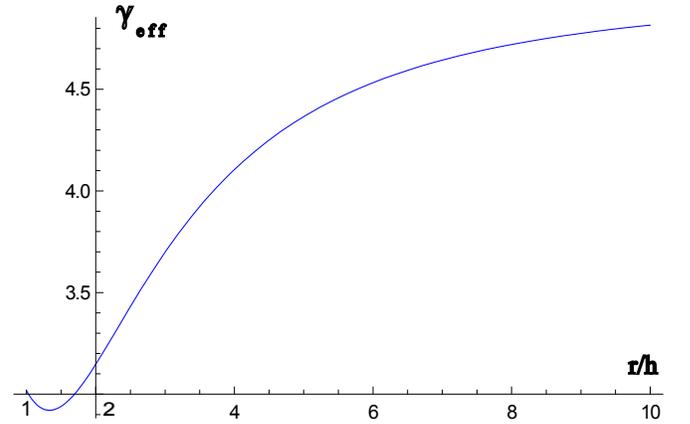}
\caption{\label{allpower}(Color online) Effective power of repulsion between two particles at the same height $h$ near the homeotropic wall in (\ref{gamma}) vs. dimensionless distance $r/h$. Corresponds to the sum of all interactions (\ref{uhsame}) for $\alpha=2,\beta=0.5$ and radius of particles $a_{1}=a_{2}=0.5h$.}
\end{figure}

Formulas (\ref{uddh})-(\ref{uqqh}) can be simply treated, as well, using \textit{image} interpretation. It is clearly seen that the interaction between two particles 1 and 2 near the homeotropic wall consists of the usual direct interaction $U_{12}$ and \textit{image} interaction between particle 1 and image of the particle 2 ( and vice-versa $2\Leftrightarrow 1$; image of the particle $p$ has dipole moment $-p$ and image of $p'$ has dipole moment $-p'$; quadrupole moments of the images are equal to quadrupole moments of real particles. See Fig.\ref{2hom}). As each of two  \textit{image} interaction has weight factor $1/2$ we come to the formulas (\ref{uddh})-(\ref{uqqh}).

If both particles are located at the same distance $h$ below the wall then $\theta=\frac{\pi}{2}$. Consider the same orientation of dipoles $p=\alpha a_{1}^{2},p'=\alpha a_{2}^{2},c=-\beta a_{1}^{3}, c'=-\beta a_{2}^{3}$. Then total energy of interaction between them:
\begin{widetext}
\begin{equation}
\frac{U_{same,hom}^{wall}}{4\pi Ka_{1}^{2}a_{2}^{2}}=\alpha^{2}\left(\frac{1}{r^{3}}+\frac{8h^{2}-r^{2}}{(4h^{2}+r^{2})^{5/2}}\right)+6\alpha\beta(a_{1}+a_{2})\frac{(3r^{2}-h^{2})h}{(4h^{2}+r^{2})^{7/2}}+\label{uhsame}
\end{equation}
\[
+3\beta^{2}a_{1}a_{2}\left(\frac{3}{r^{5}}+\frac{3}{(4h^{2}+r^{2})^{5/2}}-\frac{120h^{2}}{(4h^{2}+r^{2})^{7/2}}+\frac{560h^{4}}{(4h^{2}+r^{2})^{9/2}}  \right)\] 
\end{widetext}

This potential is repulsive, elsewhere. Let us present it, approximately, in the form of power law dependence with some effective power that depends on the distance, i.e.:

\begin{equation}
U_{same,hom}^{wall}\approx \frac{C}{r^{\gamma_{eff}}}\label{gamma}
\end{equation}

where $\gamma_{eff}$ may be found as $\gamma_{eff}=-\frac{\partial log U}{\partial log r}=-U'_{r}\frac{r}{U}$. Fig.\ref{2power} shows dependence of such effective power on the dimensionless distance $r/h$ for dipole-dipole ($\alpha\neq0, \beta=0$, see red line 1 on the Fig.\ref{2power} ) and  quadrupole-quadrupole ($\alpha=0,\beta\neq 0$, see black line 2 on the Fig.\ref{2power}). We see that on small distances $r/h \ll 1$ effective powers are 3 and 5, respectively. For large distances there is some crossover from 3 to 5 for the \textit{dipole-dipole} interaction, with the value near 4 for medium distances. This is similar to electrostatics, where, similarly, dipole interaction is weakened from the third to the fifth power near the conducting grounded wall. This power $\gamma_{eff}\approx4$ was obtained, as well, in computer simulations in \cite{andr1}.

 Quadrupole interaction has the same fifth power for very large distances $r/h \gg 1$ but in the middle area $2<r/h<5$ changes greatly with effective power being $\gamma_{eff}\approx 5.9$ for $r/h=2.5$ and $\gamma_{eff}\approx 3.8$ for $r/h=5$. 
As real particles with hedgehog configuration have both dipole and quadrupole moment, we show on the Fig.\ref{allpower} dependence of the effective power on the dimensionless distance $r/h$ for the real case $\alpha=2, \beta=0.5$ and particle radius $a=0.5h$ (all lengths should be in terms of the height \textit{h}).

\begin{figure}[ht!]
\includegraphics[clip=,width=\linewidth]{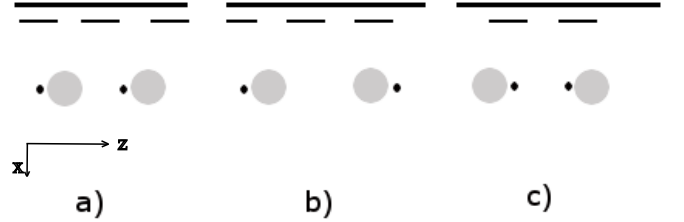}
\caption{\label{2plan}Different configurations of particles near planar wall. a) Usual dipole-dipole, b) \textit{p-p} configuration, c) \textit{h-h} configuration according to \cite{jap}.}
\end{figure}

\section{Interaction between two particles near one planar wall}
In the planar cell axis $z||\textbf{n}_{\infty}$ is parallel to the rubbing, so that the height of the particles is denoted as $x\in(0,\infty)$ (see Fig. \ref{2plan}). 
Using the Green function (\ref{gplan}) for this case, we obtain necessary potentials as shown, herein, before:

\begin{widetext}
\begin{equation}
\frac{U_{dd,plan}^{wall}}{4\pi K}=\frac{pp'}{r^{3}}(1-3cos^{2}\theta)-\frac{3pp'}{\bar{r}^{5}}(\rho^{2}sin^{2}\varphi-\bar{r}^{2}cos^{2}\bar{\theta})  \label{uddp}
\end{equation}
\begin{equation}
\frac{U_{dQ,plan}^{wall}}{4\pi K}= (pc'-cp')\frac{cos\theta}{r^{4}}(15cos^{2}\theta-9)+(pc'-cp')\frac{15\rho cos\varphi}{\bar{r}^{7}}\left( \rho^{2}sin^{2}\varphi-\bar{r}^{2}cos^{2}\bar{\theta}\right)\label{udqp}
\end{equation}
\begin{equation}
\frac{U_{QQ,plan}^{wall}}{4\pi K}= \frac{cc'}{r^{5}}(9-90cos^{2}\theta+105cos^{4}\theta)+\frac{cc'}{\bar{r}^{9}}(105\rho^{2}cos^{2}\varphi -\bar{r}^{2})(\rho^{2}sin^{2}\varphi-\bar{r}^{2}cos^{2}\bar{\theta})
\label{uqqp}
\end{equation}
\end{widetext}
Here $\rho=\sqrt{(z-z')^{2}+(y-y')^{2}}$ is the parallel projection of the distance $r$ between particles ($\rho$ lays horizontally and makes the angle $\varphi$ with $z$ direction ). As usually $\theta$ is the angle between $r$ and $z$, $$\bar{r}=\sqrt{(x+x')^{2}+(y-y')^{2}+(z-z')^{2}}$$ is the distance between particle 1 and the image of the particle 2 and $\bar{\theta}$ is the angle between $\bar{r}$ and vertical line. Then  $cos\bar{\theta}=\frac{(x+x')}{\bar{r}}>0$ and $\theta$ is the angle between $r$ and $z$. 

\subsection{Dipole-dipole interaction at the same height for the same dipole orientation, Fig.\ref{2plan}.a}

If both particles have the same height $h=h'$ then $\rho=r$ and $\varphi=\theta$. In this case the dipole-dipole energy of interaction (\ref{uddp}) between two parallel dipoles at the same height $h$ near the wall with planar anchoring conditions (Fig.\ref{2plan}.a) can be presented as:

\begin{equation}
\frac{U_{dd,plan}^{wall,same}}{4\pi Ka_{1}^{2}a_{2}^{2}}=\alpha^{2}V_{dd,plan}(\textbf{r},h)\label{udd_same_plan}
\end{equation}

\begin{equation}
V_{dd,plan}=\frac{1-3cos^{2}\theta}{r^{3}}+\frac{12h^{2}-3sin^{2}\theta r^{2}}{(4h^{2}+r^{2})^{5/2}} \label{vdd}
\end{equation}

\begin{figure}[ht!]
\includegraphics[clip=,width=\linewidth]{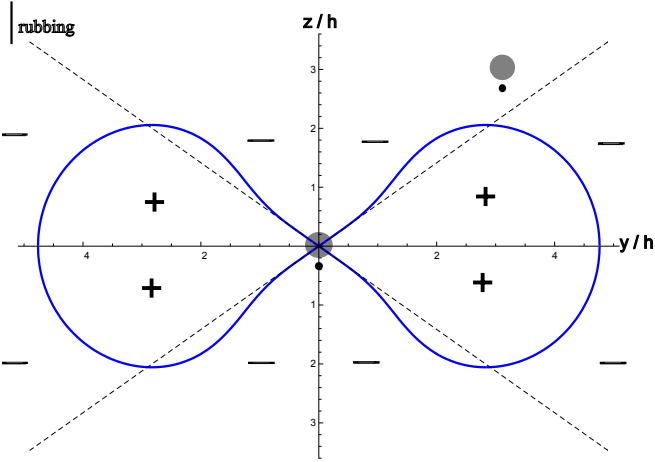}
\caption{\label{ddwallmap}(Color online) Map of the attraction and repulsion zones for dipole-dipole interaction between two particles at the same height $h$ near the planar wall. At short distances $r<h$ attraction and repulsion cones coincide with unlimited nematic case (see Fig.\ref{cones}.a) and make angle $\theta=Arccos(1/\sqrt{3})$ with $z$ axis. But at great distances $r>h$ repulsion lateral cones  dramatically collapse from the right and left side into dumbbell-shaped region. Sign '-' means attraction,'+' means repulsion (radial force $\textbf{f}_{r}=-\frac{\partial V_{dd,plan}}{\partial r}$, $\textbf{f}_{r}<0$ and $\textbf{f}_{r}>0$ respectively). On the thick blue line $\textbf{f}_{r}=0$.   This effect is absent in usual electrostatics for dipoles near grounded wall.}
\end{figure}

\begin{figure}[ht!]
\includegraphics[clip=,width=\linewidth]{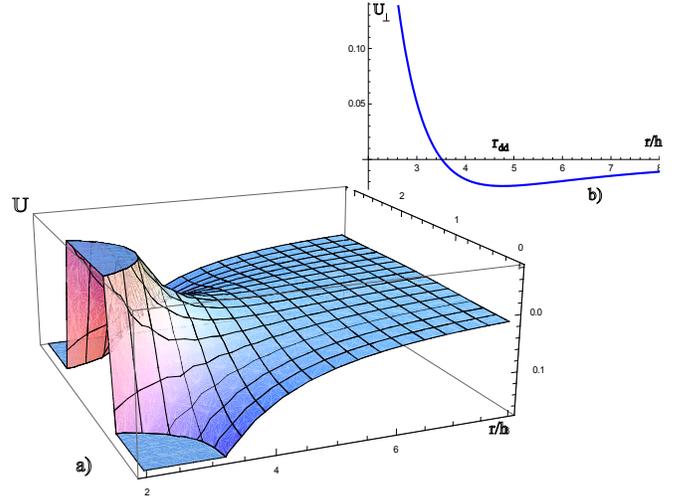}
\caption{\label{vdd1}(Color online) Dimensionless energy of dipole-dipole interaction vs. dimensionless distance \textit{r/h}. a) Energy for all angles $\theta\in (0,\pi)$, clear attraction for $\theta=0,\pi$; b) Energy $U_{\bot}$ in perpendicular direction $\theta=\pi/2$ between \textit{r} and rubbing direction \textit{z}. New effect: for $r>r_{dd}=4.76h$ \textit{attraction} appears.}
\end{figure}

\begin{figure}[ht!]
\includegraphics[clip=,width=\linewidth]{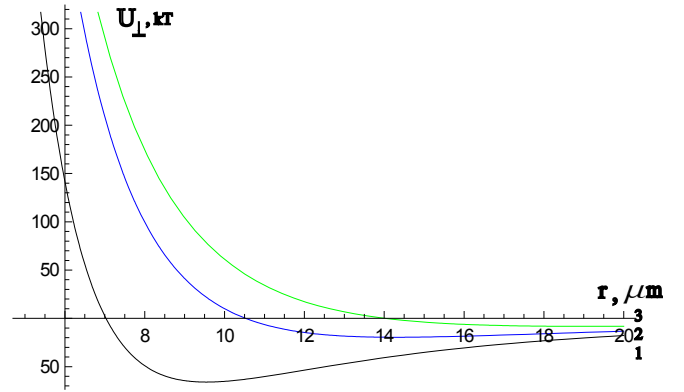}
\caption{\label{vdd2}(Color online) Energy of interaction $U_{\bot}$ in $kT$ between two dipole beads near planar wall in the perpendicular direction to the rubbing vs. distance \textit{r} in $\mu m$ for different heights from the wall: 1)\textit{h}=2 $\mu m$, 2)\textit{h}=3 $\mu m$, 3)\textit{h}=4 $\mu m$. Here $a_{1}=a_{2}=1 \mu m, \alpha=2.04, \beta=0.5$.}
\end{figure}

Let's analyze dipole-dipole interaction (\ref{vdd}) which corresponds to the usual dipole configuration on the Fig.\ref{2plan}.a. First \textit{remarkable behavior} is the dipole-dipole \textit{isotropic attraction} on far distances. Really  for \textit{any finite h}   we have asymptotic form of the dipole-dipole potential for great distances $r \gg h$
\begin{equation}
V_{dd,plan}\approx-\frac{2}{r^{3}}\label{vddasymp}
\end{equation}
 Fig.\ref{ddwallmap} represents attraction and repulsion zones obtained from the potential (\ref{vdd}). At short distances $r<h$ attraction and repulsion cones coincide with unlimited nematic case (see Fig.\ref{cones}.a) and make angle $\theta=Arccos(1/\sqrt{3})$ with $z$ axis. But at great distances $r>h$ lateral repulsion cones  dramatically collapse from the right and left side into dumbbell-shaped region. On the thick blue line radial force is zero $\textbf{f}_{r}=-\frac{\partial V_{dd,plan}}{\partial r}=0$. It crosses perpendicular direction $y/h$ in the point $r=r_{dd}=4.76h$. This means that potential $V_{dd,plan}^{\bot}$ has minimum at the distance $r_{dd}=4.76h$, so that for $r>r_{dd}$ there is attraction and for $r<r_{dd}$ there is repulsion in the perpendicular direction to the rubbing. Actually this minimum is unstable and of the saddle type as particles tends to attract along $\theta=0,\pi$ directions (see Fig.\ref{vdd1}). This effect of attraction in perpendicular direction along $\theta=\pi/2$ near the planar wall may be found only on far distances $r>r_{dd}$ where the interaction is weak. On the Fig.\ref{vdd2} we show the total potential $U_{\bot}$ in $kT$ units for the beads with radius $a=1 \mu m$ which carry dipole and quadrupole moments $\alpha=2.04, \beta=0.5$ on the distances between them in $\mu m$. Beads are located near the wall on three different heights $h=2 \mu m, h=3 \mu m,h=4 \mu m $. It is clearly seen that effect can be measurable only for $h=2-3 \mu m$. Effect of dipole attraction on far distances near the planar wall was observed as well in \cite{andr1} with help of computer similations.
 
 Actually we can say that potential between two dipole particles is approximately becomes isotropically attractive (\ref{vddasymp}) only for distances $r>r_{dd}$. This effect is absent in usual electrostatics for dipoles near grounded wall.

\subsection{Dipole-quadrupole interaction at the same height for the same dipole orientation, Fig.\ref{2plan}.a}

If both particles have the same height $h=h'$ then $\rho=r$ and $\varphi=\theta$ and we can find dipole-quadrupole potential from (\ref{udqp}). Then it can be presented in the form:

\begin{equation}
\frac{U_{dQ,plan}^{wall,same}}{4\pi Ka_{1}^{2}a_{2}^{2}}=\alpha\beta(a_{1}-a_{2})V_{dQ,plan}(\textbf{r},h) \label{udq_same_plan}
\end{equation}

\begin{equation}
V_{dQ,plan}= \frac{cos\theta(15cos^{2}\theta-9)}{r^{4}}+\frac{15cos\theta r(sin^{2}\theta r^{2}-4h^{2})}{(4h^{2}+r^{2})^{7/2}} 
\end{equation}

\begin{figure}[ht!]
\includegraphics[clip=,width=\linewidth]{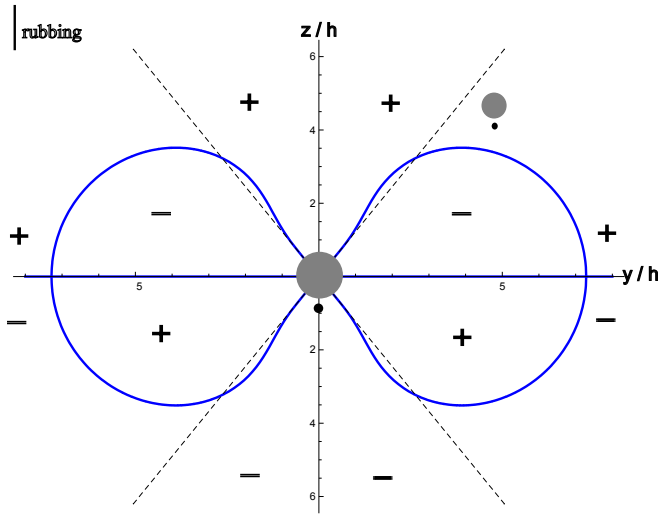}
\caption{\label{dqwallmap}(Color online) Map of the attraction and repulsion zones for dipole-quadrupole interaction (\ref{udq_same_plan}) between two particles at the same height $h$ near the planar wall with greater particle in the center.}
\end{figure}
Map of the attraction and repulsion zones of this potential is presented on the Fig.\ref{dqwallmap} (we consider greater particle is in the center $a_{1}>a_{2}$). At short distances $r<h$ attraction and repulsion cones coincide with unlimited nematic case (see Fig.\ref{cones}.b) and make angle $\theta=Arccos(3/\sqrt{15})$ with $z$ axis. But at great distances $r>h$ lateral attraction and repulsion cones dramatically collapse from the right and left side into dumbbell-shaped region. Sign '-' means attraction,'+' means repulsion (radial force $\textbf{f}_{r}=-\frac{\partial V_{dQ,plan}}{\partial r}$, $\textbf{f}_{r}<0$ and $\textbf{f}_{r}>0$ respectively). On the thick blue line $\textbf{f}_{r}=0$. It crosses axis $y/h$ in the point $r=r_{dq}=7.29h$ so that at great distances $r>r_{dq}$ this potential can be approximately presented in the form:
\begin{equation}
V_{dQ,plan}\approx\frac{6cos\theta}{r^{4}}\label{vdqasymp}
\end{equation}
which is repulsive elsewhere in the upper half-plane $z>0$ and is attractive elsewhere in the lower half-plane $z<0$. 
This effect as well is absent in usual electrostatics for dipoles and quadrupoles near grounded wall.

\subsection{Quadrupole-quadrupole interaction at the same height for the same dipole orientation, Fig.\ref{2plan}.a}

If both particles have the same height $h=h'$ then $\rho=r$ and $\varphi=\theta$ and we can find quadrupole-quadrupole potential from (\ref{uqqp}). Then it can be presented in the form:

\begin{equation}
\frac{U_{QQ,plan}^{wall,same}}{4\pi Ka_{1}^{2}a_{2}^{2}}=\beta^{2}a_{1}a_{2}V_{QQ,plan}(\textbf{r},h)\label{uqq_same_plan}
\end{equation}

\begin{equation}
V_{QQ,plan}= A_{1}+A_{2}cos^{2}\theta+A_{3}cos^{4}\theta  
\end{equation}
with 
\[
A_{1}=\frac{9}{r^{5}}+\frac{(60h^{2}-15r^{2})}{(4h^{2}+r^{2})^{7/2}},
\]
\begin{equation}
A_{2}=-\frac{90}{r^{5}}+\frac{15r^{2}}{(4h^{2}+r^{2})^{7/2}}+\frac{105r^{4}-420r^{2}h^{2}}{(4h^{2}+r^{2})^{9/2}},
\end{equation}
\[
A_{3}=\frac{105}{r^{5}}-\frac{105r^{4}}{(4h^{2}+r^{2})^{9/2}}.
\]

\begin{figure}[ht!]
\includegraphics[clip=,width=\linewidth]{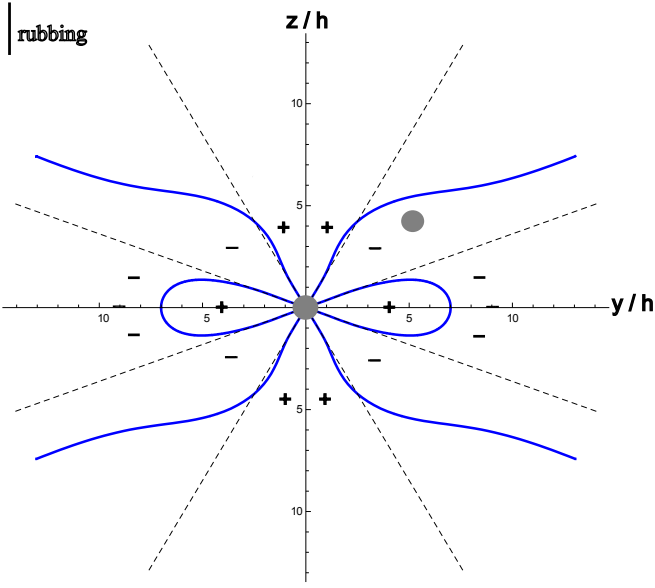}
\caption{\label{qqwallmap}(Color online) Map of the attraction and repulsion zones for quadrupole-quadrupole interaction (\ref{uqq_same_plan}) between two particles at the same height $h$ near the planar wall.}
\end{figure}
Map of the attraction and repulsion zones of this potential is presented on the Fig.\ref{qqwallmap}. At short distances $r<h$ attraction and repulsion cones coincide with unlimited nematic case (see Fig.\ref{cones}.c). But at great distances $r \gg h$ repulsion lateral cones from the right and left side  collapse into dumbbell-shaped region while upper and low cones expand from $\theta_{1}=Arccos(0.86)$ to $\theta_{1}'=Arccos(1/\sqrt{5})$ . Sign '-' means attraction,'+' means repulsion. On the thick blue line radial force $\textbf{f}_{r}=-\frac{\partial V_{QQ,plan}}{\partial r}=0$. It crosses axis $y/h$ in the point $r=r_{qq}=7.00h$ so that at great distances $r>r_{qq}$ this potential can be approximately presented in the form:
\begin{equation}
V_{QQ,plan}\approx\frac{-6+30cos^{2}\theta}{r^{5}}\label{vdqasymp}
\end{equation}
which is repulsive elsewhere in the upper and low cones with $0\leq\theta=Arccos(1/\sqrt{5})$. This effect as well is absent in usual electrostatics for quadrupoles near grounded wall.

\subsection{General elastic interaction between two particles at the same height, Fig.\ref{2plan}.a,b,c.}
 The total energy of interaction between two dipoles with parallel orientation at the same height $h$ near the wall with planar anchoring condition (Fig.\ref{2plan}.a) can be presented as:

\begin{equation}
\frac{U_{plan}^{wall,same}}{4\pi Ka_{1}^{2}a_{2}^{2}}=\alpha^{2}V_{dd,plan}(\textbf{r},h)+\alpha\beta(a_{1}-a_{2})V_{dQ,plan}(\textbf{r},h)\label{same_plan}
\end{equation}
\[
+\beta^{2}a_{1}a_{2}V_{QQ,plan}(\textbf{r},h)
\]

For untiparallel configurations we should take $p'=-p$. There are two possible cases.
Antiparallel configuration of \textit{p-p} type ($p=-p'=\alpha a^{2}$, see Fig.\ref{2plan}.b):
\begin{equation}
\frac{U_{plan}^{wall,p-p}}{4\pi Ka_{1}^{2}a_{2}^{2}}=-\alpha^{2}V_{dd,plan}(\textbf{r},h)-\alpha\beta(a_{1}+a_{2})V_{dQ,plan}(\textbf{r},h)+\label{pp}
\end{equation}
\[
+\beta^{2}a_{1}a_{2}V_{QQ,plan}(\textbf{r},h)
\]

Antiparallel configuration of \textit{h-h} type ($p=-p'=-\alpha a^{2}$, see Fig.\ref{2plan}.c):
\begin{equation}
\frac{U_{plan}^{wall,h-h}}{4\pi Ka_{1}^{2}a_{2}^{2}}=-\alpha^{2}V_{dd,plan}(\textbf{r},h)+\alpha\beta(a_{1}+a_{2})V_{dQ,plan}(\textbf{r},h)+\label{hh}
\end{equation}
\[
+\beta^{2}a_{1}a_{2}V_{QQ,plan}(\textbf{r},h)
\]
In these configurations dipole beads repel each other but in \textit{p-p} case the repulsion is weaker than in the \textit{h-h} case because of dipole-quadrupole interaction. In \textit{p-p} case there is dipole-quadrupole attraction and in \textit{h-h} case there is dipole-quadrupole repulsion.

\section{Interaction of the one particle with nematic cell}
In this section we consider two parallel walls with distance $L$ between them filled with nematic liquid crystal. In other words this is a usual nematic cell with thickness $L$. We shall consider both cases of homeotropic and planar nematic orientations in the cell.

\subsection{Interaction of the one particle with homeotropic cell}
\begin{figure}[ht!]
\includegraphics[clip=,width=\linewidth]{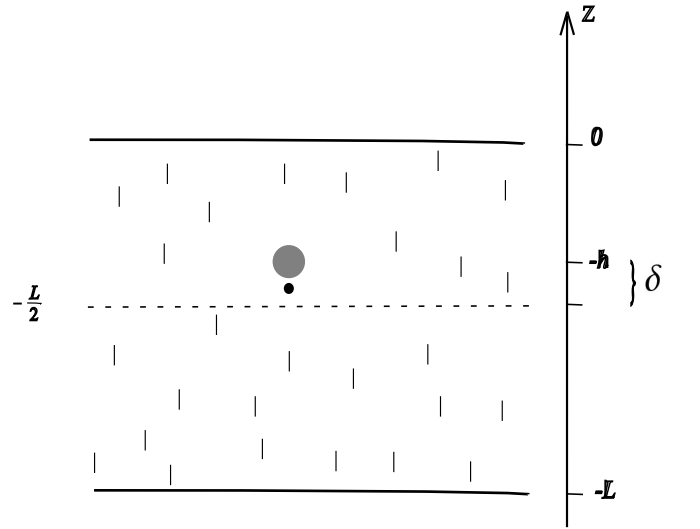}
\caption{\label{cell} One particle in the homeotropic cell. $\delta$ is the shift from the center of the cell. Equilibrium shift $\delta=\frac{\beta}{\alpha}a$ does not depend on the thickness $L$ of the cell (\ref{shift}). See Fig.\ref{1p_cell_hom} as well. }
\end{figure}

Consider first one particle in the nematic homeotropic cell. Let $h$ be the distance from the particle to the top of the cell. We choose $z=0$ at the upper plane of the cell, $z>0$ above and $z<0$ below so that particle has $z=-h$ and bottom of the cell has $z=-L$.(see Fig.\ref{cell}).

 If we consider all images of the unit charge located in the point $z=-h$ then we can construct Green function:

\begin{equation}
G_{hom}^{cell}(\textbf{x},\textbf{x}')=\frac{1}{|\textbf{x}-\textbf{x}'|}+H_{hom}^{cell}(\textbf{x},\textbf{x}') \label{ghomcell}
\end{equation}

\[
H_{hom}^{cell}(\textbf{x},\textbf{x}')=-\sum^{\infty}_{n=-\infty}\frac{1}{|\textbf{A}_{n}(\textbf{x})-\textbf{x}'|}+\sum^{\infty}_{n=-\infty \atop n\neq0}\frac{1}{|\textbf{B}_{n}(\textbf{x})-\textbf{x}'|} 
\]

where $\textbf{A}_{n}(\textbf{x})=(x,y,-z+2nL)$ and $\textbf{B}_{n}(\textbf{x})=(x,y,z+2nL)$ for $\textbf{x}=(x,y,z)$. 
Substituting $H_{hom}^{cell}$ into (\ref{uself}) brings energy of the one particle within the homeotropic cell $ U_{self}^{hom,cell}=U_{dd,self}^{hom,cell}+U_{dQ,self}^{hom,cell}+U_{QQ,self}^{hom,cell}$:
\begin{equation}
U_{dd,hom}^{self,cell}=\frac{\pi K \alpha^{2}a^{4}}{2}\sum^{\infty}_{n=-\infty}\frac{1}{|nL+h|^{3}}-\frac{\pi K \alpha^{2}a^{4}}{L^{3}}\zeta(3)\label{hdds}
\end{equation}
\begin{equation}
U_{dQ,hom}^{self,cell}=\mp\frac{3\pi K \alpha\beta a^{5}}{2}\sum^{\infty}_{n=-\infty}\frac{sign(nL+h)}{|nL+h|^{4}}\label{hdqs}
\end{equation}
\begin{equation}
U_{QQ,hom}^{self,cell}=\frac{3\pi K \beta^{2}a^{6}}{2}\sum^{\infty}_{n=-\infty}\frac{1}{|nL+h|^{5}}+\frac{3\pi K \beta^{2}a^{6}}{L^{5}}\zeta(5)\label{hqqs}
\end{equation}
where $\zeta(t)=\sum^{\infty}_{n=1}\frac{1}{n^{t}}$ is Riemann zeta function, $\zeta(3)=1.202,\zeta(5)=1.036$. 

Here $h$ is the distance from the particle to the upper plane of the cell, '-' corresponds to \textit{p}-configuration and '+' corresponds to \textit{h}-configuration, see Fig.\ref{onehom}. We see that dipole-quadrupole energy (\ref{hdqs}) is zero when particle is located in the middle of the cell $h=\frac{L}{2}$. It should be zero as well from the symmetry reasons as in the middle of the cell there is no difference between configurations with $p<0$ and $p>0$. 

Self energy of the single quadrupole particle within the cell was obtained as well in the paper \cite{fukuda1} using the \textit{coat} approach (but authors there did not find dipole-dipole and dipole-quadrupole interaction with the cell). Authors received formula similar to (\ref{hqqs}) but with unknown multiplier $\Gamma$ and without the constant $\frac{3\pi K \beta^{2}a^{6}}{L^{5}}\zeta(5)$. If we set $\Gamma=2\pi Kc=-2\beta\pi Ka^{3}$ in \cite{fukuda1} we receive the same formula (\ref{hqqs}) up to a constant.

\begin{figure}[ht!]
\includegraphics[clip=,width=\linewidth]{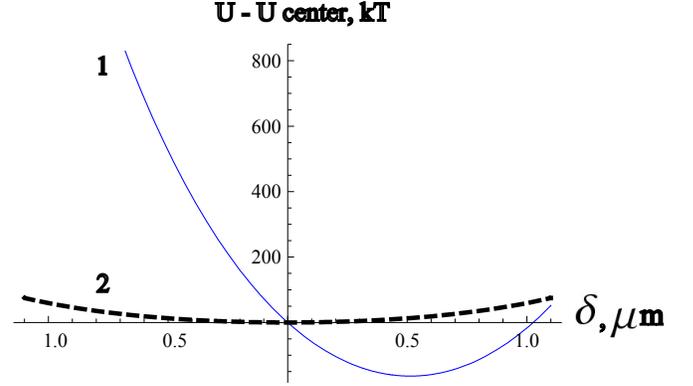}
\caption{\label{1p_cell_hom}(Color online) One particle with radius $a_{1}=2\mu m$ in the homeotropic cell with thickness $L=10\mu m$. Dependence of the energy shift $U-U_{z=L/2}$ on the shift $\delta$ from the center of the cell. Blue line 1 corresponds to the dipole particle $\alpha=2,\beta=0.5$. Equilibrium shift $\delta=0.5$ is in line with (\ref{shift}). Thick dashed black line 2 corresponds to the quadrupole particle $\alpha=0,\beta=0.5$ .}
\end{figure}

From this expressions we can obtain approximate energy of the one particle in the homeotropic nematic cell near the center of the cell if we consider first terms of series with $n=0,-1$. Then for one particle which carries dipole moment $\alpha\neq 0$ it's energy in the homeotropic nematic cell is:
\begin{equation}
U_{dd,hom}^{self,cell}\approx 4\pi K \alpha^{2}a^{4}\left(\frac{1}{(L+2\delta)^{3}}+\frac{1}{(L-2\delta)^{3}}\right)\pm\label{udd_cell}
\end{equation}
\[
\pm24\pi K \alpha\beta a^{5}\left(\frac{1}{(L+2\delta)^{4}}-\frac{1}{(L-2\delta)^{4}}\right)
\]
Then minimum condition $\frac{\partial U_{dd,hom}^{self,cell}}{\partial \delta}=0$ brings equilibrium shift of the particle from the center (see Fig.\ref{1p_cell_hom}):
\begin{equation}
\delta_{eq}=\pm\frac{\beta}{\alpha} a\label{shift}
\end{equation}
where  '+' corresponds to \textit{p}-configuration, '-' corresponds to \textit{h}-configuration (see Fig.\ref{onehom}). This means that this shift from the center of the cell is positive (up) for \textit{p}-configuration and negative (down) for \textit{h}-configuration.
Surprisingly that $\delta_{eq}$ does not depend on the thickness $L$ of the cell!

If dipole moment is zero, then energy of the one quadrupole particle in the homeotropic nematic cell is approximately:
\begin{equation}
U_{qq,hom}^{self,cell}\approx 48\pi K \beta^{2}a^{6}\left(\frac{1}{(L+2\delta)^{5}}+\frac{1}{(L-2\delta)^{5}}\right)\label{uqq_cell}
\end{equation}
where $\delta$ is the shift of the particle from the center of the cell $z=\frac{L}{2}$ ($\delta_{eq}=0$ in equilibrium).

\subsection{Interaction of the one particle with planar cell}

In this subsection we consider one colloidal particle within the nematic cell with director parallel to the planes. Let the rubbing direction be along $z||\textbf{n}_{\infty}$. We choose coordinate system for this planar cell with axis $x$ looks down $0\leq x\leq L$ and $L$ is width of the cell  (see Fig.\ref{1p_cell_plan} ) so that particle has the height $x=h$ from the upper plane. 

\begin{figure}[ht!]
\includegraphics[clip=,width=\linewidth]{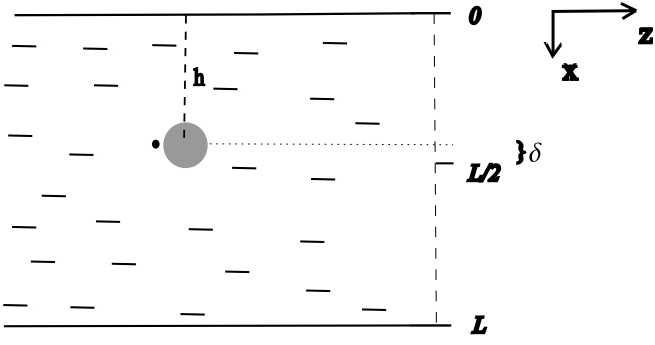}
\caption{\label{1p_cell_plan} One particle in the planar cell. $\delta$ is the shift from the center of the cell. Equilibrium shift $\delta=0$. See Fig.\ref{1p_cell_plan} as well.}
\end{figure}

Using the same procedure as in the Sec.III,B (see formula (\ref{gplan})) we construct Green function for the planar cell from the Green function (\ref{ghomcell}) for the homeotropic cell:

\begin{equation}
G_{plan}^{cell}(\textbf{x},\textbf{x}')=\frac{1}{|\textbf{x}-\textbf{x}'|}+H_{plan}^{cell}(\textbf{x},\textbf{x}') \label{gplancell}
\end{equation}

\[
H_{plan}^{cell}(\textbf{x},\textbf{x}')=-\sum^{\infty}_{n=-\infty}\frac{1}{|\tilde{\textbf{A}}_{n}(\textbf{x})-\textbf{x}'|}+\sum^{\infty}_{n=-\infty \atop n\neq0}\frac{1}{|\tilde{\textbf{B}}_{n}(\textbf{x})-\textbf{x}'|} 
\]

where $\tilde{\textbf{A}}_{n}(\textbf{x})=(-x+2nL,y,z)$ and $\tilde{\textbf{B}}_{n}(\textbf{x})=(x+2nL,y,z)$ for $\textbf{x}=(x,y,z)$.  
 
Substitution $H_{plan}^{cell}$ into (\ref{uself}) brings energy of the one particle within the planar nematic cell $ U_{self}^{plan,cell}=U_{dd,self}^{plan,cell}+U_{dQ,self}^{plan,cell}+U_{QQ,self}^{plan,cell}$:
\begin{equation}
U_{dd,plan}^{self,cell}=\frac{3\pi K \alpha^{2}a^{4}}{4}\sum^{\infty}_{n=-\infty}\frac{1}{|nL+h|^{3}}+\frac{\pi K \alpha^{2}a^{4}}{2L^{3}}\zeta(3)\label{pdds}
\end{equation}
\begin{equation}
U_{dQ,plan}^{self,cell}=0\label{pdqs}
\end{equation}
\begin{equation}
U_{QQ,plan}^{self,cell}=\frac{15\pi K \beta^{2}a^{6}}{16}\sum^{\infty}_{n=-\infty}\frac{1}{|nL+h|^{5}}+\frac{9\pi K \beta^{2}a^{6}}{8L^{5}}\zeta(5)\label{pqqs}
\end{equation}

Self energy of the single quadrupole particle within the cell was obtained as well in the paper \cite{fukuda1} (but authors there did not find dipole-dipole and dipole-quadrupole interaction with the cell). Authors received formula similar to (\ref{pqqs}) but with unknown multiplier $\Gamma$ and without the constant $\frac{9\pi K \beta^{2}a^{6}}{8L^{5}}\zeta(5)$. If we set $\Gamma=-2\beta\pi Ka^{3}$ in \cite{fukuda1} we receive the same formula (\ref{pqqs}) up to a constant.

Expressions (\ref{pdds}-\ref{pqqs}) show that the equilibrium position of the particle within the planar cell is the center of the cell $z=\frac{L}{2}$. We can obtain similarly approximate energy of the one particle in the planar nematic cell near the center of the cell if we consider first terms of series with $n=0,-1$. Then for one particle which carries dipole moment it's energy in the planar cell:
\begin{equation}
U_{dd,plan}^{self,cell}\approx 6\pi K \alpha^{2}a^{4}\left(\frac{1}{(L+2\delta)^{3}}+\frac{1}{(L-2\delta)^{3}}\right)\label{uddp_cell}
\end{equation}
If dipole moment is zero, then energy of the one quadrupole particle in the planar nematic cell is approximately:
\begin{equation}
U_{qq,plan}^{self,cell}\approx 30\pi K \beta^{2}a^{6}\left(\frac{1}{(L+2\delta)^{5}}+\frac{1}{(L-2\delta)^{5}}\right)\label{uqqp_cell}
\end{equation}
where $\delta$ similar is the shift of the particle from the center of the cell $z=\frac{L}{2}$.

In the paper \cite{lavr3} authors used formula (\ref{udd_cell}) obtained for homeotropic cell to find dependence of the shift $\delta$ from the cell thickness $L$ in the \textit{planar cell}. This is not correct though they received some qualitative agreement with experimental data. In fact there formula (\ref{uddp_cell}) for the planar cell should be used to fit experimental results more accurately.

\begin{figure}[ht!]
\includegraphics[clip=,width=\linewidth]{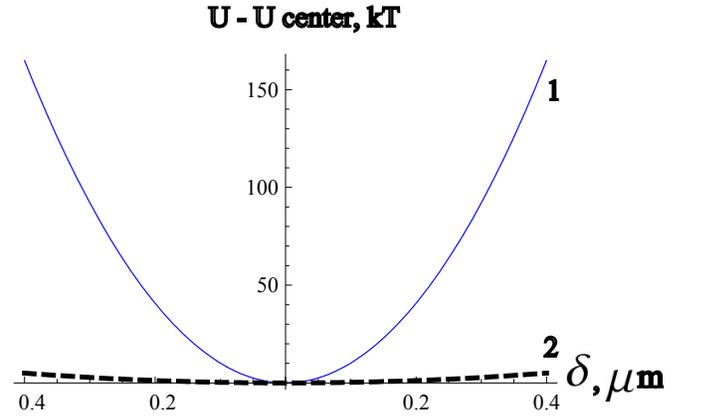}
\caption{\label{1p_cell_plan}(Color online) One particle with radius $a_{1}=2\mu m$ in the planar cell with thickness $L=10\mu m$. Dependence of the energy shift $U-U_{z=L/2}$ on the shift $\delta$ from the center of the cell. Blue line 1 corresponds to the dipole particle $\alpha=2,\beta=0.5$. Thick dashed black line 2 corresponds to the quadrupole particle $\alpha=0,\beta=0.5$ .} 
\end{figure}

\section{Interaction between two particles within the homeotropic nematic cell}
\begin{figure}[ht!]
\includegraphics[clip=,width=\linewidth]{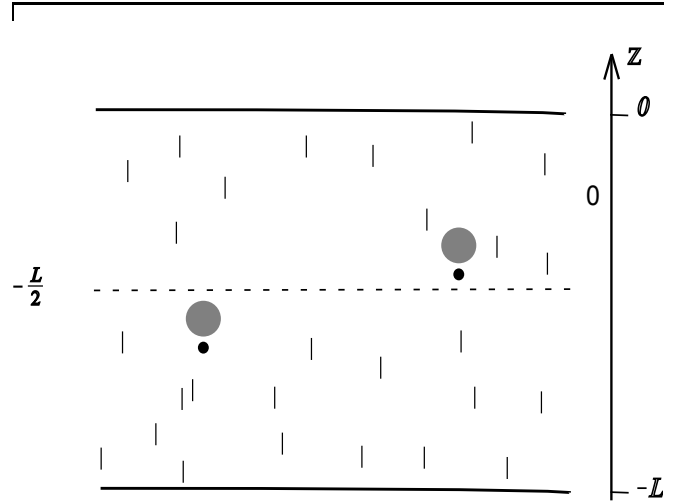}
\caption{\label{2phomcell} Two particles in the homeotropic cell with thickness $L$.}
\end{figure}

In this section we shall consider interaction between two particles located within the nematic cell with homeotropic alignment on both walls.
 The thickness of the cell is supposed to be $L$ (see Fig (\ref{2phomcell})). 
 
 Actually we already have Green function (\ref{ghomcell}) for this case. But it is much more convenient to use another form of this Green function which is commonly used in electrodynamics (see \cite{jac}):
\begin{widetext}
\begin{equation}
G_{hom}^{cell}(\textbf{x},\textbf{x}')=\frac{4}{L}\cdot\sum_{n=1}^{\infty}\sum_{m=-\infty}^{\infty}e^{im({\varphi-\varphi'})}sin\frac{n\pi z}{L}sin\frac{n\pi z'}{L}I_{m}(\frac{n\pi \rho_{<}}{L})K_{m}(\frac{n\pi \rho_{>}}{L})\label{g1}
\end{equation}
\end{widetext}

\begin{figure}[ht!]
\includegraphics[clip=,width=\linewidth]{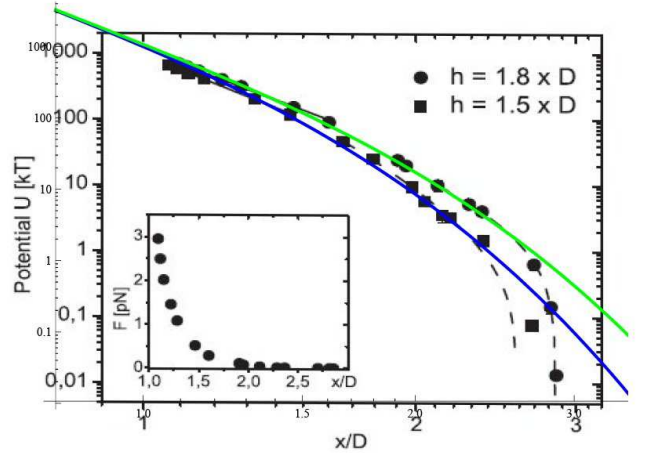}
\caption{\label{qqexp}(Color online)  Experimental data taken from \cite{conf} - energy of elastic interaction between two spherical particles with diameter $D=2a=4.4\mu m$ in the homeotropic cell with thicknesses $h=6.5\mu m$ and $h=8\mu m$. Solid blue and green lines are taken from the formula (\ref{qqhom}). The only unknown parameter $\beta=0.28$ fits both thicknesses very well in the energy scale $1 \div 1000kT$.}
\end{figure}

Here heights $z,z'$, horizontal projections $\rho_{<},\rho_{>}$ and $I_{m},K_{m}$ are modified Bessel functions.
Then using of (\ref{uint}) brings dipole-dipole interaction in the cell :
\begin{equation}
U_{dd,cell}^{hom}=\frac{16\pi K pp'}{L^{3}}\sum_{n=1}^{\infty}(n\pi)^{2} sin\frac{n\pi z}{L}sin\frac{n\pi z'}{L}K_{0}(\frac{n\pi \rho}{L})\label{dhom}
\end{equation}
Dipole-quadrupole interaction is:
\begin{equation}
U_{dQ,cell}^{hom}=\frac{16\pi K}{L^{4}}\sum_{n=1}^{\infty}(n\pi)^{3}K_{0}(\frac{n\pi \rho}{L})\times\label{dqhom}
\end{equation}
\[
\times\left[pc'\cdot sin\frac{n\pi z}{L}cos\frac{n\pi z'}{L}+p'c \cdot cos\frac{n\pi z}{L}sin\frac{n\pi z'}{L}\right]
\]
We see that if particles are located in the middle of the cell $z=z'=\frac{L}{2}$ then dipole-quadrupole interaction is zero $U_{dQ,cell}^{hom}=0$.
Similar quadrupole-quadrupole interaction takes the form:
\begin{equation}
U_{QQ,cell}^{hom}=\frac{16\pi K cc'}{L^{5}}\sum_{n=1}^{\infty}(n\pi)^{4}cos\frac{n\pi z}{L}cos\frac{n\pi z'}{L}K_{0}(\frac{n\pi \rho}{L})\label{qqhom}
\end{equation}
with $\rho$ being the horizontal projection of the distance between particles. 
Fig.\ref{qqexp} demonstrates application of this formula (\ref{qqhom}) for the repulsion potential between two spherical particles (with planar anchoring on the surface providing quadrupole director configuration $\alpha=0, \beta\neq0$) with diameter $D=2a=4.4\mu m$ in the center of homeotropic cell ($z=L/2$) with thicknesses $L=h=6.5\mu m$ and $L=h=8\mu m$ ( experimental data are taken from \cite{conf} ). The only unknown parameter $\beta=0.28$ ( we remind that $c=-\beta a^{3}$ ) fits both thicknesses very well in the energy scale $1 \div 1000kT$. Spherical particles in this experiment \cite{conf} have boojum director configuration (see Fig.\ref{defects}c.) and there is no exact estimations for $\beta$ for this case. But the value $\beta=0.28$ is very reasonable and is in accordance with experimental estimation from \cite{noel}, where authors found $\beta=0.2$ for hedgehog configuration Fig.\ref{defects} a.

The similar formula (\ref{qqhom}) was obtained in the paper \cite{fukuda} within the \textit{coat} approach with unknown multiplier $\Gamma$ as the parameters of the coat remain unknown and it is difficult to estimate them correctly for the case of strong anchoring condition $Wa/K\geq1$. If we set $\Gamma=2\pi Kc=-2\beta\pi Ka^{3}$ in \cite{fukuda} we receive the same formula (\ref{qqhom}).


\section{Interaction between two particles within the planar nematic cell}

\begin{figure}[ht!]
\includegraphics[clip=,width=\linewidth]{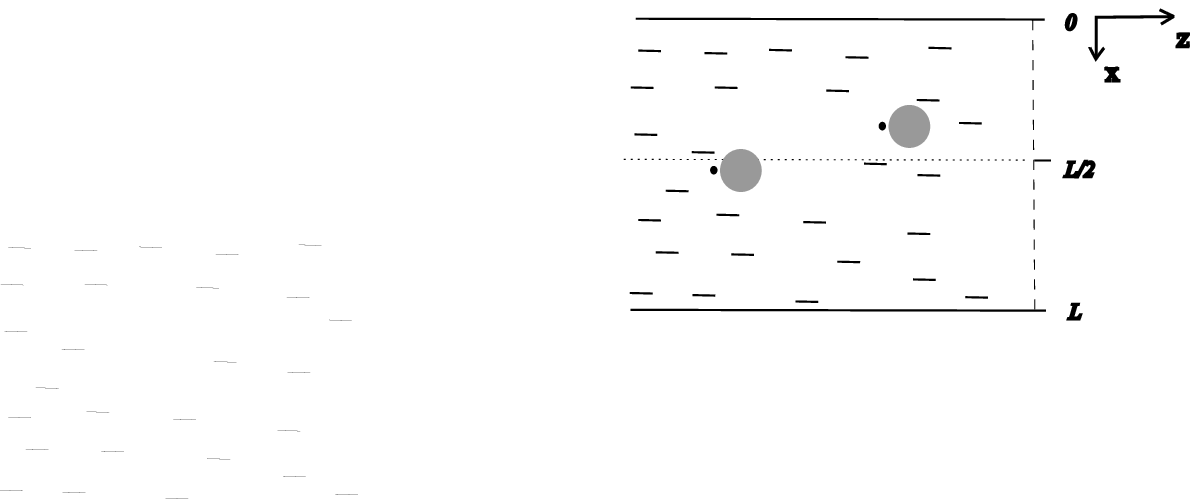}
\caption{\label{2pplancell}Two particles in the planar cell with thickness $L$. }
\end{figure}
In this section we shall consider interaction between two particles located within the nematic cell with planar alignment on both walls. The thickness of the cell is supposed to be $L$  ( see Fig \ref{2pplancell} ).

 Although we already have Green function (\ref{gplancell}) for this case it is much more convenient to use another form of this Green function which can be obtained from (\ref{g1}) with help of coordinate system transformation like in Sec III,B. Let's turn coordinate system (CS) of the homeotropic cell $CS^{hom}$ $(x,y,z)$ round the $y$ axis on $\pi/2$. Then we will have $CS^{plan}$ $(\tilde{x},\tilde{y},\tilde{z})$ with transition matrix $A$: $\textbf{x}=A\tilde{\textbf{x}},\textbf{x}'=A\tilde{\textbf{x}}'$ so that $x=\tilde{z},y=\tilde{y},z=-\tilde{x}$. Then $G_{hom}(\textbf{x},\textbf{x}')=G_{hom}(A\tilde{\textbf{x}},A\tilde{\textbf{x}}')=G_{plan}(\tilde{\textbf{x}},\tilde{\textbf{x}}')$. Omitting sign $\sim$ we may write Green function for planar cell in the $CS^{plan}$ with $\textbf{n}||z$ and $x$ perpendicular to the cell plane 
 ($x\in [0,L]$):
\begin{widetext}
\begin{equation}
G_{plan}^{cell}(\textbf{x},\textbf{x}')=\frac{4}{L}\cdot\sum_{n=1}^{\infty}\sum_{m=-\infty}^{\infty}e^{im({\varphi-\varphi'})}sin\frac{n\pi x}{L}sin\frac{n\pi x'}{L}I_{m}(\frac{n\pi \rho_{<}}{L})K_{m}(\frac{n\pi \rho_{>}}{L})\label{g2}
\end{equation}
where heights of particles $x,x'$, horizontal projections $\rho_{<}=\sqrt{y^{2}+z^{2}},\rho_{>}=\sqrt{y'^{2}+z'^{2}}$ , $tg \varphi = \frac{y}{z},tg \varphi' = \frac{y'}{z'}$ and $\rho_{<}$ is less than $\rho_{>}$.
Then taking derivatives from (\ref{uint}) brings all interactions in the planar cell. Dipole-dipole interaction:
\begin{equation}
U_{dd,cell}^{plan}=\frac{16\pi K pp'}{L^{3}}(F_{1}-F_{2}cos^{2}\varphi)\label{uplanfi}
\end{equation}
where
$$
F_{1}=\sum_{n=1}^{\infty}\frac{(n\pi)^{2}}{2}sin\frac{n\pi x}{L}sin\frac{n\pi x'}{L}\left[K_{0}(\frac{n\pi \rho}{L})+K_{2}(\frac{n\pi \rho}{L})\right]-(n\pi)^{2}cos\frac{n\pi x}{L}cos\frac{n\pi x'}{L}K_{0}(\frac{n\pi \rho}{L}),
$$
$$
F_{2}=\sum_{n=1}^{\infty}(n\pi)^{2}sin\frac{n\pi x}{L}sin\frac{n\pi x'}{L}K_{2}(\frac{n\pi \rho}{L}) .
$$
\end{widetext}

\begin{figure}[ht!]
\includegraphics[clip=,width=\linewidth]{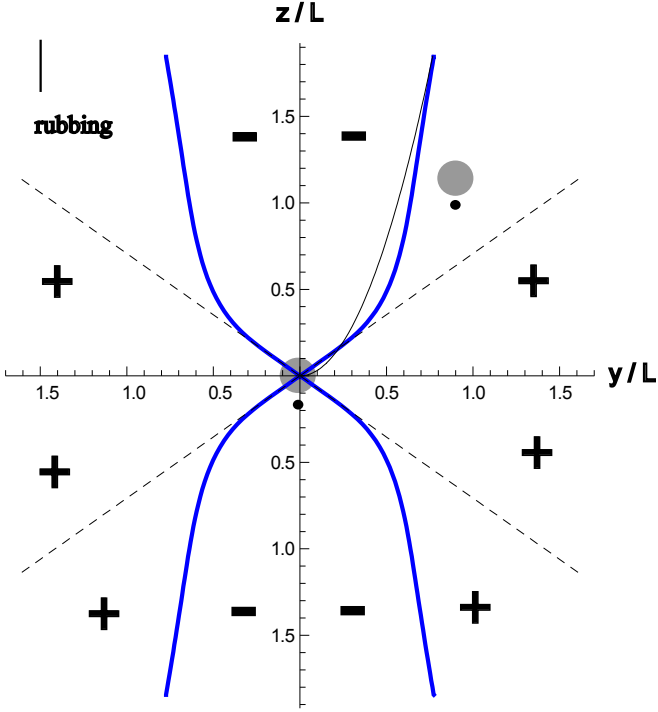}
\caption{\label{ddcellmap}(Color online) Map of the attraction and repulsion zones for dipole-dipole interaction (\ref{uplanfi}) between two particles at the center of the planar cell with thickness $L$. Director $\textbf{n}_{0}||z$. At short distances attraction and repulsion cones coincide with unlimited nematic case (see Fig.\ref{cones}.a) and make angle $\theta=Arccos(1/\sqrt{3})$ with $z$ axis. But at great distances $r\geq L$ attraction cones are deformed to the parabola $z=\frac{\pi y^{2}}{L}$.  Black thin line is the parabola $z=\frac{\pi y^{2}}{L}$. Sign '-' means attraction,'+' means repulsion. On the thick blue line radial force $\textbf{f}_{r}=-\frac{\partial V_{dd,cell}^{plan}}{\partial r}=0$. }
\end{figure}
with $\rho$ being the horizontal projection of the distance between particles and $\varphi$ is the azimuthal angle between $\rho$ and $z$.

Dipole-quadrupole interaction:
\begin{equation}
U_{dQ,cell}^{plan}=\frac{16\pi K}{L^{4}}(pc'-cp')cos\varphi(C_{1}+C_{2}cos^{2}\varphi)\label{udqplanfi}
\end{equation}
where
$$
C_{1}=L\left( F_{1\rho}'-\frac{2F_{2}}{\rho}\right),
$$
$$
C_{2}=L\left( \frac{2F_{2}}{\rho}-F_{2\rho}'\right),
$$

\begin{figure}[ht!]
\includegraphics[clip=,width=\linewidth]{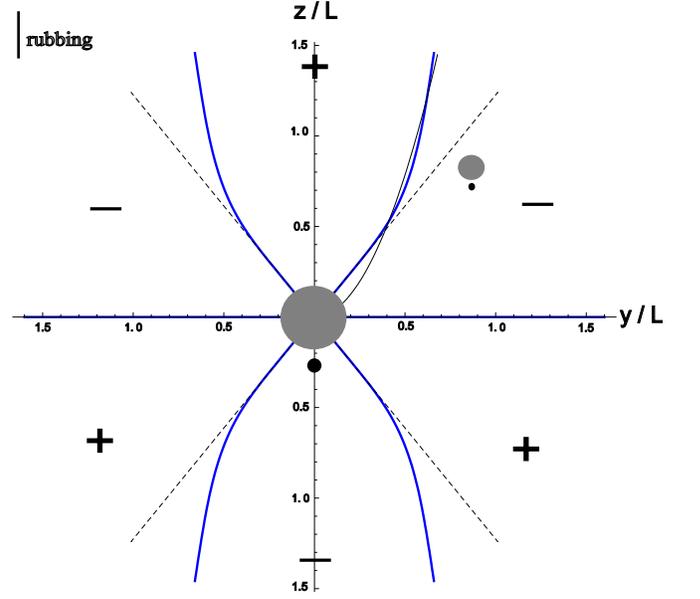}
\caption{\label{dqcellmap}(Color online) Map of the attraction and repulsion zones for dipole-quadrupole interaction (\ref{udqplanfi}) between two particles at the center of the planar cell with thickness $L$ with greater particle in the center. Black thin line is the parabola $z=\frac{\pi y^{2}}{L}$.}
\end{figure}

Quadrupole-quadrupole interaction:
\begin{equation}
U_{QQ,cell}^{plan}=\frac{16\pi K cc'}{L^{5}}(D_{1}+D_{2}cos^{2}\varphi+D_{3}cos^{4}\varphi)\label{uqqplanfi}
\end{equation}
where
$$
D_{1}=L^{2}\left(\frac{2F_{2}}{\rho^{2}}-\frac{F_{1\rho}'}{\rho}\right),
$$
$$
D_{2}=L^{2}\left( -\frac{10F_{2}}{\rho^{2}}+\frac{5F_{2\rho}'}{\rho}+\frac{F_{1\rho}'}{\rho} -F_{1\rho\rho}''\right),
$$
$$
D_{3}=L^{2}\left( \frac{8F_{2}}{\rho^{2}}-\frac{5F_{2\rho}'}{\rho}+F_{2\rho\rho}''\right),
$$

\begin{figure}[ht!]
\includegraphics[clip=,width=\linewidth]{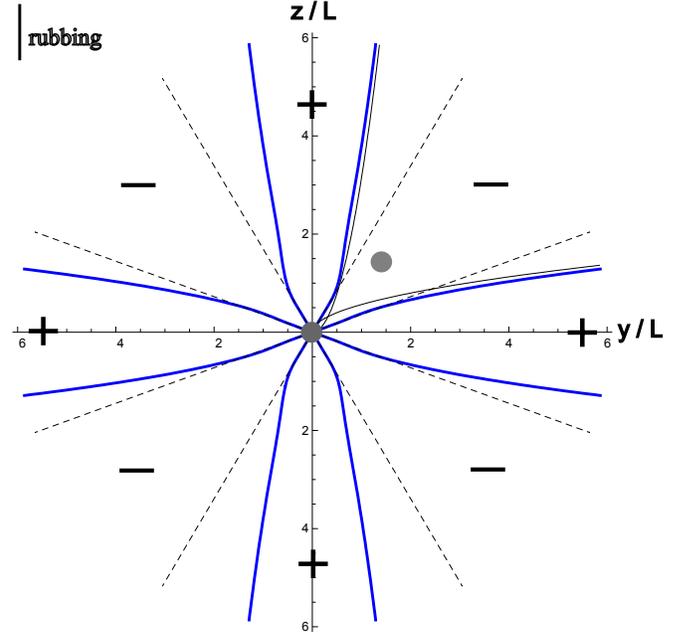}
\caption{\label{qqcellmap}(Color online) Map of the attraction and repulsion zones for quadrupole-quadrupole interaction (\ref{uqqplanfi}) between two particles at the center of the planar cell with thickness $L$. Black thin lines are the parabolas $z=\frac{\pi y^{2}}{L}$ and $y=\frac{\pi z^{2}}{L}$ respectively.}
\end{figure}

When both particle are located in the center of the cell $x=x'=\frac{L}{2}$ we have $F_{1}=\sum_{n=1,odd}^{\infty}\frac{(n\pi)^{2}}{2}\left[K_{0}(\frac{n\pi \rho}{L})+K_{2}(\frac{n\pi \rho}{L})\right]$ $ - \sum_{n=2,even}^{\infty}(n\pi)^{2}K_{0}(\frac{n\pi \rho}{L})$ and $F_{2}=\sum_{n=1,odd}^{\infty}(n\pi)^{2}K_{2}(\frac{n\pi \rho}{L})$ ($F_{1\rho}'<0,F_{2\rho}'<0$). In the limit of small distance $\rho \ll L$ between particles these functions have asymptotics $F_{1}\rightarrow \frac{L^{3}}{4\rho^{3}}$ and $F_{2}\rightarrow \frac{3L^{3}}{4\rho^{3}}$ so that we come to the well known result $U_{dd,cell}^{plan}=\frac{4\pi K pp'}{\rho^{3}}(1-3cos^{2}\varphi)$ for $\rho \ll L$. 
In the limit of big distances $\rho \geq L$ we have $\frac{F_{1\rho}'}{F_{2\rho}'}=1-\frac{L}{\pi \rho}+o(\frac{L}{\rho})$ with accuracy $5\%$ already for $\rho=L$. So for $\rho \geq L$ the radial component of the force between particles may be written as $\textbf{f}_{\rho} = -\frac{\partial U_{dd,plan}^{c}}{\partial \rho}=-\frac{16\pi K pp'}{L^{3}}F_{2\rho}'(\rho)\cdot(1-\frac{L}{\pi \rho}-cos^{2}\varphi)$ so that dipole-dipole interaction is attractive ($\textbf{f}_{\rho}<0$) for $-\varphi_{c}\leq\varphi<\varphi_{c}$, $\varphi_{c}=arccos(\sqrt{1-\frac{L}{\pi \rho}})\approx \sqrt{\frac{L}{\pi \rho}}$ and is repulsive for $\varphi_{c}<\varphi<2\pi-\varphi_{c}$ (if dipoles are parallel each other $p=p'$ and vice versa if $p=-p'$  ). In other words for $\rho>L$ dipole-dipole interaction is attractive inside parabola $z=\frac{\pi y^{2}}{L}$ and is repulsive outside this parabola (see Fig.\ref{ddcellmap}).

Similarly in the limit of small distance $\rho \ll L$ between particles we come to the result of unlimited nematic $U_{dQ,cell}^{plan}=\frac{4\pi K}{\rho^{4}}(pc'-cp')cos \varphi(15cos^{2}\varphi-9)$ and $U_{QQ,cell}^{plan}=\frac{4\pi Kcc'}{\rho^{5}}(9-90cos^{2}\varphi+105cos^{4}\varphi)$ for $\rho \ll L$. 

Map of the attraction and repulsion zones of the dipole-quadrupole potential is presented on the Fig.\ref{dqcellmap} (we consider greater particle is in the center $a_{1}>a_{2}$). At short distances $r<0.5L$ attraction and repulsion cones coincide with unlimited nematic case (see Fig.\ref{cones}.b) and make angle $\theta=Arccos(3/\sqrt{15})$ with $z$ axis. But at great distances $r>0.5L$ upper and low cones shrink to the parabola $z=\frac{\pi y^{2}}{L}$. 

Map of the attraction and repulsion zones of the quadrupole-quadrupole potential is presented on the Fig.\ref{qqcellmap}. At short distances $r<L$ attraction and repulsion cones coincide with unlimited nematic case (see Fig.\ref{cones}.c). But at great distances $r \gg L$ upper and low repulsion cones shrink to the parabola $z=\frac{\pi y^{2}}{L}$ and lateral repulsion cones shrink to the parabola $y=\frac{\pi z^{2}}{L}$.

\section{Discussion}

Here we want to discuss the current approach in the context of others theoretical approaches used for description of colloidal particles in NLC. We will emphasize initial assumptions of them as well as relationships between them.

1. \textbf{\textsl{The current approach}} (\cite{lupe,we} and this paper). Initial assumptions.

Let one particle creates director field on far distances $n_{\mu}(\textbf{r})=p\frac{\mu}{r^{3}}+3c\frac{\mu z}{r^{5}}$ with $\mu=x,y$. This means that only things that other particles \textit{'see'} and \textit{'feel'} are these two constants $p$ and $c$ through the director field. Therefore as initial points of this approch we take these two constants \textit{plus principle of least free energy}. So this approach can be called: \textit{constants plus principle of least free energy}. Nothing else.

Then it enables to introduce some effective  functional $F_{eff}$ (\ref{flin}) which gives necessary solutions on the extremal of this functional. And this potential gives all necessary potentials between particles as a consequence (\ref{uself}), (\ref{uint}).

2. \textbf{\textsl{Coat approach}} \cite{lev,lev3,fukuda1,fukuda}. This approach was suggested by authors in \cite{lev3}. 

Initial assumptions of the coat approach are the following: some volume of the NLC around the particle which contains all strong deformations and defects is called the \textit{coat}. Outside the coat deformations are small. On the surface of the coat some anchoring coefficient  $W(\sigma)$ \textit{is artificially introduced }(1st assumption) which depends on the point $\sigma$ of the coat so that surface energy takes the form:
\begin{equation}
F_{s}=\oint_{coat} d\sigma W(\sigma)(\textbf{n}\nu )^{2}\label{fs}
\end{equation}
Symmetry of the $W(\sigma)$ coincides with the symmetry of the director field around the particle. Plus possibility of the gradient expansion of the director is supposed (2nd assumption) $a\partial n_{\mu}\ll1$. So the coat approach =  \textit{gradient expansion of the director $a\partial n_{\mu}\ll1$ plus unknown anchoring  $W(\sigma)$}. In the case of small particles without defects the coat coincides with the particle itself $W(\sigma)=W_{surface}$. But even in this case one assumption is necessary: coat approach =\textit{gradient expansion of the director $a\partial n_{\mu}\ll1$}.

Then it enables to find interaction potentials for any form of the coat and to find the connection between the symmetry and the type of the interaction potential \cite{lev3}. As well this approach was used for description of the spherical particle in confined NLC - in homeotropic and planar nematic cell \cite{fukuda1,fukuda}. Surprisingly this approach gave the same results as in formulas (\ref{3p}), (\ref{hqqs}) and (\ref{qqhom}) but with unknown multiplier $\Gamma$ ! If we set $\Gamma=2\pi Kc=-2\beta\pi Ka^{3}$ in \cite{fukuda1,fukuda} we receive the same formulas (\ref{3p}), (\ref{hqqs}) up to a constant and (\ref{qqhom}).

But gradient expansion demand $a\partial n_{\mu}\ll1$ can not be satisfied quantitatively on the surface of the particle as  director is changed on the distances comparable to the size $a$ of the particle. Really for dipole case $n_{\mu}(\textbf{r})=p\frac{\mu}{r^{3}}$ we have $a\partial n_{\mu}\approx ap/a^{3}\sim \alpha \approx 1$ near the particle. Therefor the coat approach can not give exact quantitative results. Nevertheless it is very surprising that it gives similar results as (\ref{3p}), (\ref{hqqs}) and (\ref{qqhom}). 

Apparently the reason of such resemblance is the following: in the coat approach \textit{another effective functional is obtained as intermediate result} after gradient expansion on the coat surface. Actually there surface term takes the form up to a constant \cite{lev3}:
\begin{equation}
F_{s}=\alpha_{3\mu}n_{\mu}+\beta_{3s\nu}\partial_{s}n_{\mu}+\gamma_{3sk\mu}\partial_{s}\partial_{k}n_{\mu}
\end{equation}
with constants $\alpha_{3\mu},\beta_{3s\nu},\gamma_{3sk\mu}$ taken as surface integrals on the coat. They play the role of charge, dipole and quadrupole moments. So that total effective functional obtained there has the form:
\begin{widetext}
\begin{equation}
F_{coat}=K\int d^{3}x\left\{\frac{(\nabla n_{\mu})^{2}}{2}+\alpha_{3\mu}n_{\mu}+\beta_{3s\nu}\partial_{s}n_{\mu}+\gamma_{3sk\mu}\partial_{s}\partial_{k}n_{\mu}\right\} \label{ours}
\end{equation}
\end{widetext}
And now it is apparent that quantitative estimates of these constants $\alpha_{3\mu},\beta_{3s\nu},\gamma_{3sk\mu}$ obtained in \cite{lev3} are wrong \textit{though the form of the functional is true} (as we think). In the case of axially symmetric particles we have $\alpha_{3\mu}=0$ and the functional takes the form (\ref{flin}). It is necessary to abandon simple connection between constants and the surface and \textit{to consider constants as just constants}. As far as this connection is not trivial and can not be obtained within the bounds of suggestion $a\partial n_{\mu}\ll1$. It takes to make experiment or to make computer simulation to find these constants. Then approaches 1. and 2. are in fact equivalent. 

3. \textbf{\textsl{Approach of \cite{perg2,perg3}.}} 

In this approach the particle is surrounded by imaginary sphere which contains all defects inside and \textit{the director is supposed to be fixed firmly on the surface of the sphere}. Outside the sphere director follows Laplace equation. Thus this fixed director on the surface plays the role of the source of deformations in the bulk NLC. Then Green function on the sphere enables to find director field in the whole space and to find interaction potentials between particles. So authors do not make gradient expansion as in \cite{lev3}.

Qualitatively results of this approach conforms with two previous approaches. Similarly interaction potential between particles was obtained as the multipole series expansion \cite{perg2} and even three particles effects have been found. This approach was applied as well for the interaction between one dipole particle and the wall. Comparison of (\ref{1}) and (\ref{1p}) shows that repulsion of the one particle with dipole moment from the planar wall is 3/2 times stronger than from the homeotropic wall  $U_{dd,self}^{plan,wall}/U_{dd,self}^{hom,wall}=3/2$. Authors of \cite{perg3} as well obtained this ratio to be 3/2. 

Nevertheless this approach differs quantitatively from the previous ones. Approach of \cite{perg2,perg3} predicts dipole-dipole force to be three times weaker and quadrupole-quadrupole five times weaker than results of \cite{lupe} \textit{for the same director field on far distances}. This discrepancy has no roots in different definitions for dipole or quadrupole moment. If someone consider director field in \cite{perg2} near the particle and rewrite it in the form $n_{\mu}(\textbf{r})=p\frac{\mu}{r^{3}}+3c\frac{\mu z}{r^{5}}$ with $\mu=x,y$ and then compare elastic interaction between two particles in the same definitions $p$ and $c$ then dipole-dipole force will be 3 times weaker and quadrupole-quadrupole force will be 5 times weaker than results of \cite{lupe} for unbounded NLC.

Actually this discrepancy is unclear now.

All three approaches contain some \textit{additional assumptions} for the simplification of the problem. No one of them solves the problem in the exact formulation described in Sec.II . 

Until now all experimental efforts about colloidal particles in nematic liquid crystals were concentrated on the study of interaction potentials and structures of the particles only \cite{po1}-\cite{conf}. But experiments on the precise measurement of the director field \textit{near one particle} are absent absolutely. Exactly this kind of experiments which can measure precisely director field and elastic interaction between particles \textit{simultaneously} can be a testing area for validity of the one approach or another.

\section{Conclusion}

To conclude we apply method developed in \cite{we} for theoretical investigation of colloidal elastic interactions between axially symmetric particles in the confined nematic liquid crystal (NLC) near one wall and in the nematic cell with thickness $L$. Both cases of homeotropic and planar director orientations are considered. Particularly dipole-dipole, dipole-quadrupole and quadrupole-quadrupole interactions of the \textit{one} particle with the wall and within the nematic cell are found as well as correspondent \textit{two particle} elastic interactions.
Set of new results has been predicted:  effective power of repulsion between two dipole particles at height $h$ near the homeotropic wall is reduced gradually from inverse 3 to 5 with increasing of dimensionless distance $r/h$; near the planar wall the effect of dipole-dipole \textit{isotropic attraction} is predicted on the large distances $r>r_{dd}=4.76 h$; maps of attraction and repulsion zones are crucially changed for all interactions near the planar wall and in the planar cell; one dipole particle in the homeotropic nematic cell was found to be shifted on the distance $\delta_{eq}$ from the center of the cell \textit{independent} of the thickness $L$ of the cell. 
The proposed theory fits very well experimental data for the confinement effect of elastic interaction between spheres in the homeotropic cell taken from \cite{conf} in the range $1\div1000 kT$.

These surface induced effects show the differences between nematostatics and electrostatics which can be tested experimentally. 
All numerical calculations in the paper were performed using Mathematica 6 and in all series we used summation $\sum_{n=1}^{100}$.

\end{document}